\documentclass[12pt,letterpaper]{article}%
\usepackage{amsmath,amssymb,amsfonts}
\usepackage[pdftex]{graphicx}
\usepackage{float}
\usepackage{caption2}
\usepackage[]{hyperref}
\usepackage[Symbolsmallscale]{upgreek}
\usepackage{dsfont}
\usepackage{graphicx}

\numberwithin{equation}{section}
\setcaptionmargin{1cm}
\newcommand{\hhref}[1]{\href{http://arxiv.org/abs/#1}{arXiv:#1}}
\begin{document}
\begin{titlepage}
\begin{flushright}
\end{flushright}
\vskip 1.0cm
\begin{center}
{\Large \bf Radiation Problem in Transplanckian Scattering} \vskip
1.0cm {\large Paolo Lodone and
Slava Rychkov} \\[0.7cm]
{\it Scuola Normale Superiore and INFN, Pisa, Italy} \vskip 2.0cm
\end{center}
\begin{abstract}
We investigate hard radiation emission in small-angle transplanckian
scattering. We show how to reduce this problem to a quantum field
theory computation in a classical background (gravitational shock
wave). In momentum space, the formalism is similar to the flat-space
light cone perturbation theory, with shock wave crossing vertices
added. In the impact parameter representation, the radiating
particle splits into a multi-particle virtual state, whose
wavefunction is then multiplied by individual eikonal factors.

As a
phenomenological application, we study QCD radiation in
transplanckian collisions of TeV-scale gravity models. We derive the
distribution of initial state radiation gluons, and find a
suppression at large transverse momenta with respect to the standard
QCD result. This is due to rescattering events, in which the quark
and the emitted gluon scatter coherently. Interestingly, the
suppression factor depends on the number of extra dimensions and
provides a new experimental handle to measure this number. We
evaluate the leading-log corrections to partonic cross-sections due
to the initial state radiation, and prove that they can be absorbed
into the hadronic PDF. The factorization scale should then be chosen
in agreement with an earlier proposal of Emparan, Masip, and
Rattazzi.

In the future, our methods can be applied to the gravitational
radiation in transplanckian scattering, where they can go beyond the
existing approaches limited to the soft radiation case.
\end{abstract}
\vskip 1.5cm September 2009
\end{titlepage}

\section{Introduction}

Scattering at center-of-mass (CM) energies exceeding the quantum gravity scale
(transplanckian scattering, or \textit{T-scattering}, for short) is an exotic
process of significant theoretical interest. In particular, it provides a
laboratory to study the black hole information loss paradox. Microscopic black
hole formation and its subsequent evaporation is expected for impact
parameters $b$ of the order of the Schwarzschild radius $R_{S}$ of a black
hole of mass $\sqrt{s}$ \cite{tHooft},\cite{BF},\cite{EG},\cite{GR}. The
detailed description of how this happens depends on the unknown underlying
theory of quantum gravity and is at present out of reach. On the other hand,
large impact parameters $b\gg R_{S}$ correspond to elastic small-angle
scattering, whose amplitude can be predicted on the basis of General
Relativity alone. It is given by eikonalized single-graviton exchange
\cite{tHooft},\cite{ACV87},\cite{Verlindes},\cite{Kabat:1992tb}. Computing the
corrections in $b/R_{S}$ to the elastic scattering, one hopes to learn about
the strong inelastic dynamics at $b\sim R_{S}$ \cite{Amati:1987uf}%
,\cite{Amati:1992zb},\cite{Ciafaloni:2008dg},\cite{Veneziano:2008xa},\cite{Giddings}.

\begin{figure}[h]
\begin{center}
\includegraphics[
height=1.5in
]{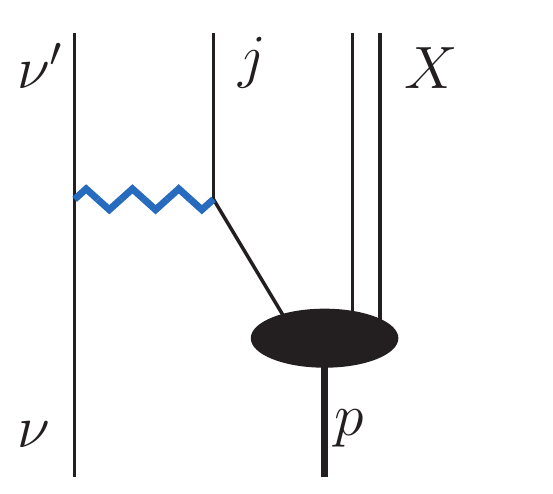}\hspace{1in}\includegraphics[ height=1.5in
]{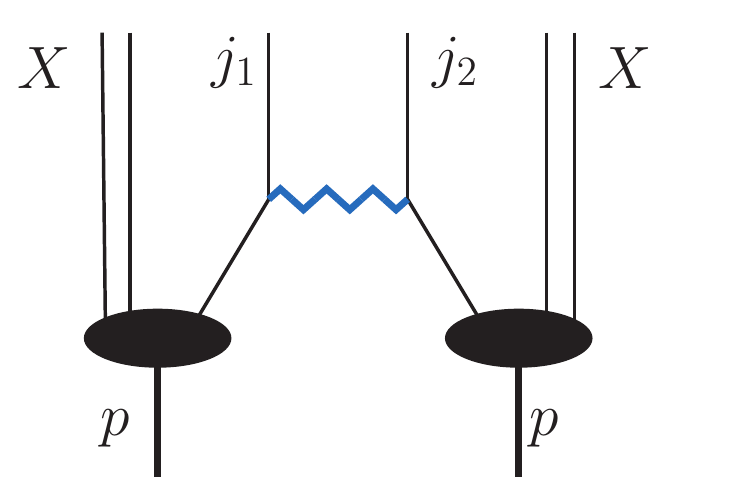}
\end{center}
\caption{\textit{The $\nu p$ T-scattering in cosmic ray physics (left) and the
$pp$ T-scattering at the LHC (right). In both cases the dominant process is
small-angle elastic scattering between partons, giving jet(s)$+$anything in
the final state. The zigzag line denotes the eikonal $2\to2$ scattering
amplitude.}}%
\label{fig:DISpp}%
\end{figure}

T-scattering is also interesting phenomenologically. If large extra dimension
scenarios of TeV-scale gravity \cite{ADD} are realized in Nature, this process
could be observed at the LHC and other future colliders \cite{factory}%
,\cite{GRW}, as well as in collisions of high-energy cosmic neutrinos with
atmospheric nucleons \cite{BHinCR},\cite{Emparan:2001kf}. In these scenarios
the total T-scattering cross section is finite, grows with energy, and is
dominated by calculable small-angle scattering between partonic constituents
\cite{Emparan:2001kf},\cite{GRW}, see Fig.\ \ref{fig:DISpp}. The subleading
black hole production cross section at present can only be estimated from
geometrical arguments.

\begin{figure}[h]
\begin{center}
\includegraphics[
height=1.5in
]{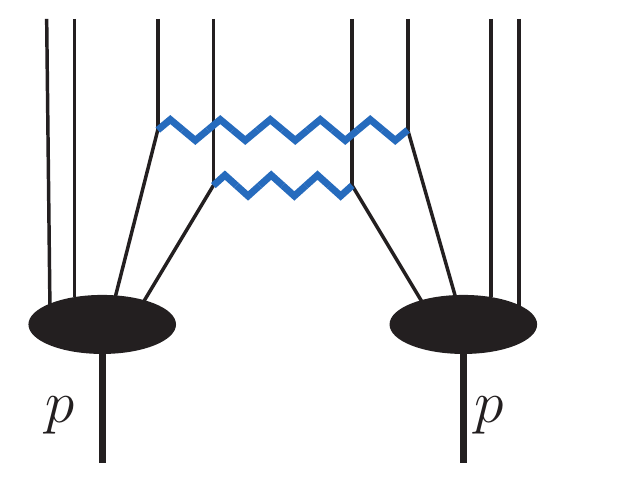}
\end{center}
\caption{\textit{Multiple partonic interactions, in which more than
one pair of partons exchange gravitons, are suppressed. This is
because the eikonal phase vanishes very quickly with the transverse
separation, and it is unlikely
to find a second pair of partons for which this phase is large.}}%
\label{fig:pp}%
\end{figure}

In spite of the small scattering angle, the typical momentum transfer in these
scattering events is well above the QCD scale, and the typical impact
parameter is much smaller than the proton size, which sets the typical
distance between two uncorrelated partons inside the proton. It is unlikely
that a multiple parton interaction, Fig.~\ref{fig:pp}, will occur in the same
T-scattering. Thus it is clear that the partonic picture should be applicable
at leading order in the QCD coupling. In other words, we can compute the total
cross section via a convolution of the partonic cross section and the parton
distribution functions (PDF) $f(x,\mu_{F}^{2})$.

\begin{figure}[h]
\begin{center}
\includegraphics[
height=1.5in
]{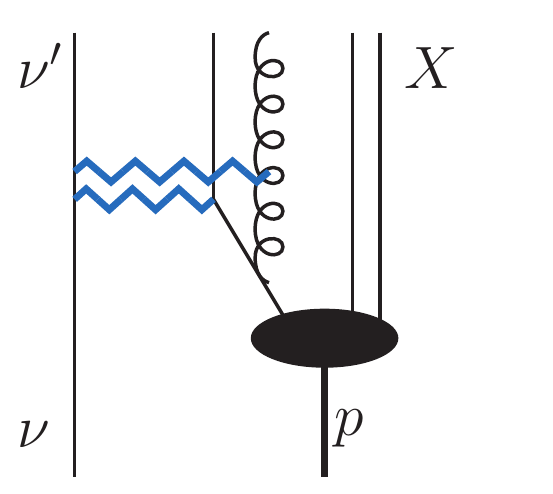}
\end{center}
\caption{\textit{Initial state radiation processes, when a quark emits a gluon
just before participating in T-scattering, are expected to play a role. They
give logarithmic corrections to the cross section, which determine the optimal
factorization scale $\mu_{F}$ of the process.}}%
\label{fig:DISg}%
\end{figure}

Several interesting questions arise when one tries to think what happens
beyond the leading order. For example, it's not known how to treat events of
the type shown in Fig.\ \ref{fig:DISg}, where one of the colliding partons
(say a quark) radiates a gluon just before the collision. Now the quark-gluon
separation is not necessarily large, and the pair may scatter coherently. What
is the correct description of such \textit{rescattering} processes, and what
is the resulting effect on the total T-scattering cross section?

A related question is: which factorization scale $\mu_{F}$ should we choose in
the computation of the total cross section? As is well known, QCD-initiated
processes have significant higher-order logarithmic corrections, associated
with the collinear QCD radiation off initial partons. By choosing $\mu_{F}$
appropriately, these corrections can and should be reabsorbed into the PDF. As
we will see below, the familiar choice $\mu_{F}^{2}\sim-t$ is likely not the
right one for the T-scattering. If so, we would like to see this explicitly.

The purpose of this paper is to answer the above-mentioned questions. We will
be focusing on the QCD radiation since it is the dominant phenomenological
effect due to the relative largeness of $\alpha_{s}$. However, our methods are
equally applicable to the radiation of photons or any other spin 1 gauge
bosons. We also hope that these methods may later prove useful in the more
complicated problem of \textit{gravitational} radiation emitted in
T-scattering, and in particular to provide an alternative to the existing
computations which are limited to the case when the emitted radiation is
\textit{soft }\cite{Amati:1987uf},\cite{Galtsov}.

The paper is organized as follows. In Section \ref{sec:review} we review the
eikonalization of small-angle $2\rightarrow2$ partonic scattering amplitude,
following \cite{Emparan:2001kf},\cite{GRW}. To keep close contact with
phenomenology, we work in the context of large extra dimension scenarios with
the quantum gravity scale around a TeV. A key feature of the extra-dimensional
situation with $n$ compactified dimensions is the appearance of a new length
scale $b_{c}$, which sets the range of a typical T-scattering interaction. In
$D$-dimensional Planck units, $D=4+n,$ we have
\[
b_{c}\sim s^{1/n},\quad R_{S}\sim s^{1/[2(n+1)]},
\]
so that $b_{c}\gg R_{S}$ in the deep transplanckian regime $\sqrt{s}\gg1$. We
also present an alternative computation of the $2\rightarrow2$ amplitude,
based on generalizing to $D$ dimensions the early idea of 't Hooft
\cite{tHooft}, who considered the small-angle T-scattering by solving the
Klein-Gordon equation for one particle propagating in the classical
gravitational field of the other particle.

In Section \ref{sec:hadronic} we start discussing small-angle T-scattering
with hadronic initial states, as in Fig.~\ref{fig:DISpp}, in which QCD effects
are expected to play a role. According to the existing proposal of Emparan,
Masip and Rattazzi \cite{Emparan:2001kf}, the total cross section for these
processes must be computed with the following prescription for the PDF scale
$\mu_{F}$ :%
\begin{equation}
\mu_{F}(q)=\left\{
\begin{array}
[c]{l}%
q\,\quad\text{if}\quad q<b_{c}^{-1}\,,\\
q^{\frac{1}{n+1}}\left( b_{c}^{-1}\right) ^{\frac{n}{n+1}}\,\text{\quad
if\quad}q>b_{c}^{-1}\,,
\end{array}
\right. \label{eq:mu}%
\end{equation}
where $q\equiv\sqrt{-t}$. For sufficiently high momentum transfers this
deviates from the familiar prescription $\mu_{F}\simeq q$. An intuitive
justification for this scale in terms of the typical impact parameter was
given in \cite{Emparan:2001kf}, but we would like to check it via a direct computation.

The first step is to be able to evaluate the amplitude for small-angle
T-scattering accompanied by collinear QCD radiation. In the resummation
approach \cite{Emparan:2001kf},\cite{GRW}, this computation seems
prohibitively difficult even for one-gluon emission. Indeed, the eikonal
amplitude for quark-quark T-scattering is a sum of an infinite number of
crossed ladder gravition exchanges. The outgoing gluon may be attached
anywhere on the quark lines, both external and internal. Moreover, in the
$q\rightarrow q+g$ splitting, the emitted near-collinear gluon is not
necessarily soft, and thus may also exchange gravitons. The number of diagrams
to resum skyrockets.

't Hooft's approach is a much better starting point. As we point out, it can
be easily `upgraded' to the case when radiation is present, provided that only
one of the two colliding particles radiates. This covers completely
lepton-quark scattering and is an important special case for quark-quark
scattering. The idea is very simple. In the $2\rightarrow2$ scattering, 't
Hooft treated one particle classically, the other one quantum mechanically.
The only new twist is to allow the quantum particle to radiate. In other
words,~we should treat the non-radiating parton classically, while the
radiating parton \textit{and the gluonic radiation field with which it
interacts} quantum mechanically.

This trick reduces the problem to a quantum field theory computation in the
classical gravitational background produced by a relativistic point particle,
the Aichelburg-Sexl (AS) shock wave \cite{AS}. In Section \ref{sec:QFT} we
develop the necessary formalism. We first consider the simplest perturbative
quantum field theory in the AS background: a scalar field with cubic
self-interactions. We introduce a diagram technique for computing arbitrary
transition amplitudes in this theory, which turns out to be closely related to
the standard rules of light-cone perturbation theory in flat space. We then
explain the changes necessary for the gluon field and for the scalar-gluon
interactions, and compute the one-gluon emission amplitude as an example.

Notice that while fermionic matter fields can be considered analogously, we do
not include them in this work in order to keep technical details to a minimum.
Thus we stick to a toy model in which the partonic constituents of colliding
hadron(s) are scalars.

Armed with the knowledge of gluon emission amplitudes, in Section
\ref{sec:QCDeffects} we attack the question of QCD corrections to
T-scattering. For definiteness and simplicity, we consider the gravitational
analogue of the DIS: a transplanckian electron-proton collision. The
observable is the total cross section as a function of the Bjorken $x$ and the
transverse momentum transfer $\mathbf{q}$, in the small-angle region
$|\mathbf{q|}\ll\sqrt{s}$. At leading order (LO) in the QCD coupling
$\alpha_{s}$, the partons scatter elastically on the electron (no gluon
emission). At next-to-leading order (NLO), we demonstrate the appearance of
logarithmic corrections whose scale is precisely the $\mu_{F}(q)$ from
Eq.~(\ref{eq:mu}). We find that the cross section factorizes, in the sense
that~these logarithms appear multiplied by the DGLAP splitting functions, and
can be reabsorbed into the PDFs. Finally, we are able to show that this
factorization holds to all orders in $\alpha_{s}$, in the leading-logarithm
approximation (LLA).

Our computation gives an explicit check for the validity of the partonic
picture for the T-scattering. Moreover, it gives an interesting and unexpected
explanation for why the PDF scale deviates from the usual $\mu_{F}\sim q$. It
turns out that rescatterings like in Fig.~\ref{fig:DISg} \textit{suppress the
initial state QCD radiation at transverse momenta} $q\gtrsim b_{c}^{-1}$. As a
result the transverse momentum distribution of emitted gluons has the form:%
\begin{equation}
\frac{dN}{d^{2}\mathbf{q}}=f(q)\left( \frac{dN}{d^{2}\mathbf{q}}\right)
_{\text{0}}\,, \label{eq:dN}%
\end{equation}
where $\left( dN/d^{2}\mathbf{q}\right) _{\text{0}}\sim1/\mathbf{q}^{2}$ is
the standard distribution without rescattering, and $f(q)$ is a function
interpolating between $1$ for $q\ll b_{c}^{-1}$ and $1/(n+1)$ for $q\gg
b_{c}^{-1}$. Logarithmic corrections to the cross section are obtained, as
usual, by integrating Eq.~(\ref{eq:dN}) over the gluon phase space, and the
scale of these logarithms is a geometric mean of $q$ and $b_{c}^{-1}$ as in
Eq.~(\ref{eq:mu}).

Notice that one could imagine other distributions giving rise to $\log\mu
_{F}(q)$, for example the standard $1/\mathbf{q}^{2}$ with a sharp cutoff at
$b_{\ast}^{-1}$. In this sense Eq.~(\ref{eq:dN}) contains more information
than the identification of the correct factorization scale. The predicted
suppression of the initial state radiation is $n$-dependent and could in
principle be used to determine the number of extra dimensions.

A crucial insight into the physics of radiative processes is obtained by going
into the impact parameter representation. In this picture, we find that the
scattering is described via a multi-particle wavefunction of the virtual state
(parton$+$radiated quanta), which is multipled by individual eikonal factors
when crossing the shock wave. This interpretation suggests a possible
generalization of our formalism to the case when both colliding partons
radiate, which we discuss in Section \ref{sec:simul}.

In conclusion, this work shows that factorization holds for QCD effects in
T-scattering, and that the factorization scale has a nontrivial dependence on
$q^{2}$, in agreement with the earlier proposal of Ref.~\cite{Emparan:2001kf}.
The novelty is that we arrive at these results by a concrete computation, and
that we derive the modified distribution of the initial state radiation due to
rescattering effects. The new distribution should be now incorporated into a
`transplanckian parton shower algorithm', to be used in Monte-Carlo
simulations of T-scattering. We will come back to this issue in a future publication.

\section{Review of the eikonal approach to T-scattering}

\label{sec:review}

In this section we review the basics of small angle T-scattering in the
eikonal approximation. We will work within the large flat extra dimensions
scenario of TeV-scale gravity \cite{ADD}, see \cite{PDG} for the current
experimental constraints.

Consider then two transplanckian massless Standard Model (SM) particles, thus
confined to the SM 3-brane, which scatter due to the $D$-dimensional
gravitational field, $D=4+n$, $n$ being the number of large extra dimensions,
$n\geq2$ for phenomenological reasons. For now we ignore all interactions
except for gravity. In particular, we suppose that the colliding particles are
not charged, and thus cannot emit photons or gluons. We are interested in the
scattering amplitude for small momentum transfer $-t/s\ll1.$ In this regime
gravitational radiation is also suppressed (see \cite{GRW}), and we have
elastic $2\rightarrow2$ scattering.

\subsection{Resummation}

\label{sec:resum}

\begin{figure}[h]
\begin{center}
\includegraphics[
height=1in ]{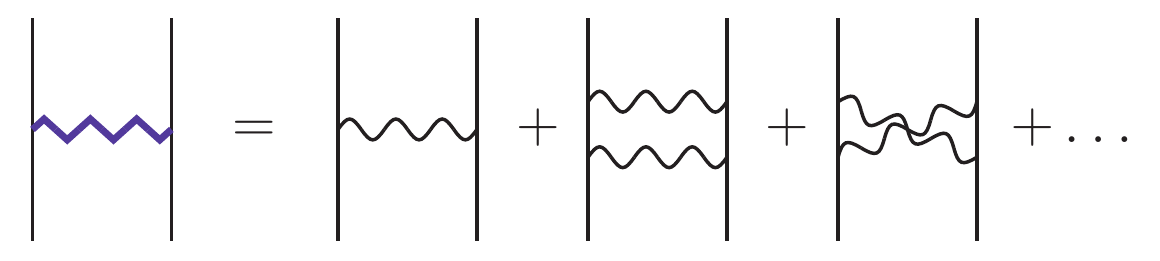}
\end{center}
\caption{\textit{The $2\to2$ small angle T-scattering amplitude is given by a
sum of crossed-ladder graviton exchanges.}}%
\label{fig:eik}%
\end{figure}

The most direct way to compute the amplitude is by resumming the crossed
ladder graviton exchange diagrams \cite{Emparan:2001kf},\cite{GRW}, see
Fig.\ \ref{fig:eik}. For small momentum transfer, exchanged gravitons are
soft, and well-known simplifications occur in the vertices and the
intermediate state propagators, allowing the resummation. The first term in
the series, the one-graviton exchange, is given by%
\begin{equation}
\mathcal{A}_{\text{Born}}(\mathbf{q})=-\frac{s^{2}}{M_{D}^{n+2}}\int
\frac{d^{n}l}{\mathbf{q}^{2}+l^{2}}\,, \label{eq:Born}%
\end{equation}
where $\mathbf{q}$ is the momentum transfer, which lies mostly in the
direction transverse to the beam: $t\approx-\mathbf{q}^{2}$. The
$D$-dimensional Planck scale $M_{D}\sim1$ TeV is normalized as in
\cite{GRW},\cite{PDG}. The divergent integral over the extra dimensional
momentum $l$ needs to be treated properly; see below. The second term in the
series, the sum of two one-loop diagrams, turns out to be equal to a
convolution of two Born amplitudes:%
\[
\mathcal{A}_{\text{1-loop}}(\mathbf{q})=\frac{i}{4s}\int\frac{d^{2}\mathbf{k}%
}{(2\pi)^{2}}A_{\text{Born}}(\mathbf{k})A_{\text{Born}}(\mathbf{q}%
-\mathbf{k})~\text{,}%
\]
and this pattern continues to higher orders. As a result the series can be
summed by going to the impact parameter representation. The amplitude acquires
the eikonal form:%
\begin{equation}
\mathcal{A}_{\text{eik}}(\mathbf{q})=\mathcal{A}_{\text{Born}}+\mathcal{A}%
_{\text{1-loop}}+\ldots=-2is\int d^{2}\mathbf{b}\,e^{-i\mathbf{q}.\mathbf{b}%
}(e^{i\chi}-1)\,, \label{eq:AeikR}%
\end{equation}
with the eikonal phase $\chi$ given by the Fourier transform of the Born
amplitude in the transverse plane:%
\[
\chi(\mathbf{b})=\frac{1}{2s}\int\frac{d^{2}\mathbf{q}}{(2\pi)^{2}%
}e^{i\mathbf{q}.\mathbf{b}}A_{\text{Born}}(\mathbf{q})\,.
\]

\begin{figure}[h]
\begin{center}
\includegraphics[
width=5.405in ]{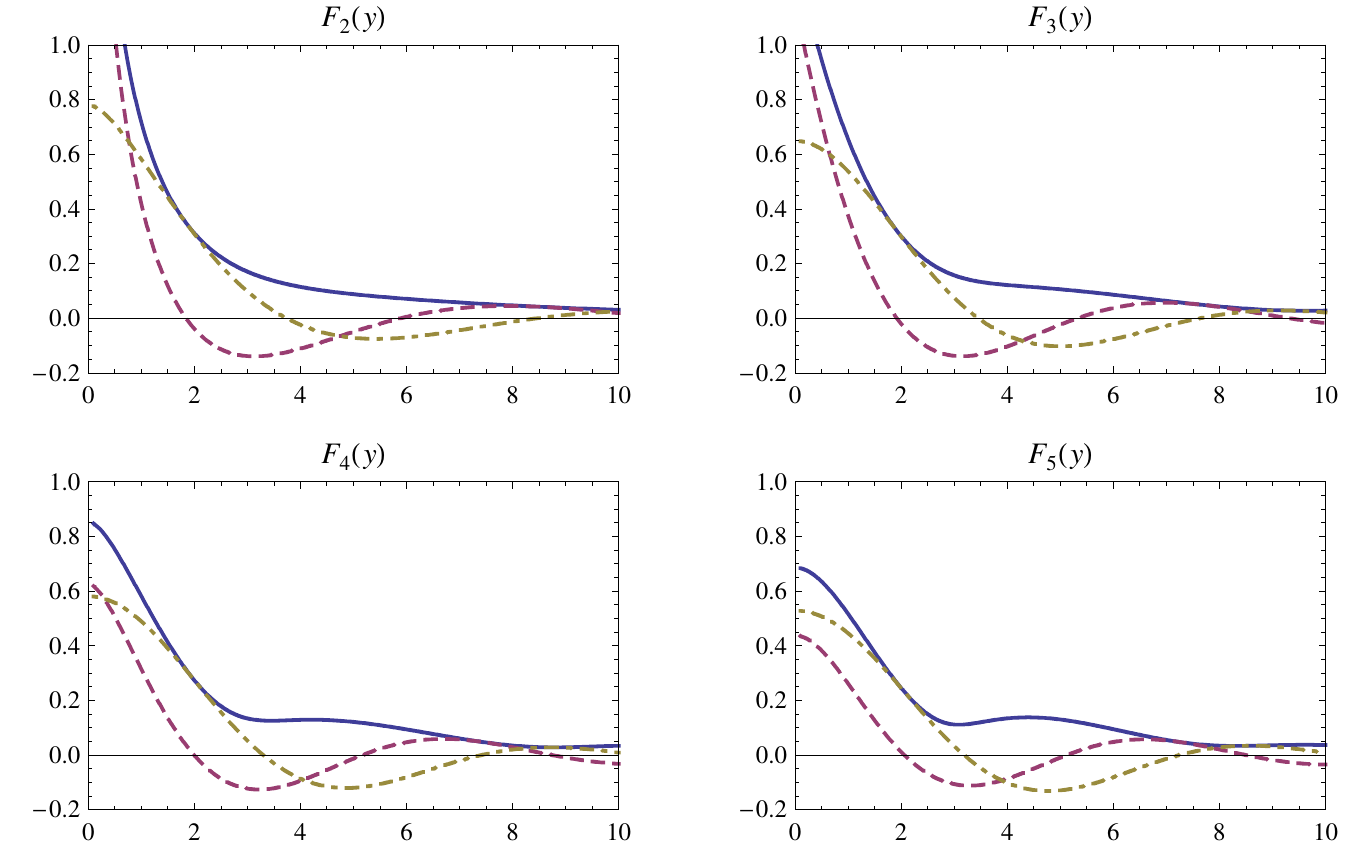}
\end{center}
\caption{\textit{The functions $F_n(y)$ for $n=2,3,4,5$: absolute
value (solid), real part (dashed), imaginary part (dot-dashed).
Notice that $\Im m\, F_n(0)<\infty$ for all $n\ge2$, implying finite
total cross section.
On the other hand, $\Re e\, F_n(0)<\infty$ for all $n\ge 3$. See \cite{GRW}.}}%
\label{fig:Fn}%
\end{figure}

To evaluate the eikonal phase, we need to regulate the divergent Born
amplitude (\ref{eq:Born}). In \cite{GRW}, dimensional regularization was used,
and it was argued that since the subtracted divergent terms are local, they do
not affect the small angle scattering amplitude\footnote{This was later
confirmed in \cite{Sjodahl:2006gb} by using in (\ref{eq:Born}) a physical
regulator $\exp(-k^{2}w^{2})$, with $w$ interpreted as an effective width of
the SM brane, $w\sim$TeV$^{-1}$. It was found that the resulting eikonal phase
coincides with (\ref{eq:eikR}) for $b\gtrsim w$, while for $b\lesssim w$ it
varies slowly (logarithmically). The eikonal amplitude then indeed agrees with
(\ref{eq:Fn}) in the small scattering angle region. The same conclusion was
also reached in \cite{Illana} using a sharp cutoff.}. The eikonal phase was
found to be:%
\begin{equation}
\chi(\mathbf{b})=\left( \frac{b_{c}}{|\mathbf{b}|}\right) ^{n},\quad
b_{c}=\frac{1}{M_{D}}\left[ \frac{(4\pi)^{\frac{n}{2}-1}\Gamma(n/2)}%
{2}\right] ^{1/n}\left( \frac{s}{M_{D}^{2}}\right) ^{1/n}\,.
\label{eq:eikR}%
\end{equation}
The corresponding amplitude is then given by:%
\begin{gather}
\mathcal{A}_{\text{eik}}=4\pi s\,b_{c}^{2}F_{n}(b_{c}|\mathbf{q}%
|)~,\nonumber\\
F_{n}(y)=-i\int_{0}^{\infty}dx\,xJ_{0}(xy)[e^{ix^{-n}}-1]\,. \label{eq:Fn}%
\end{gather}
The functions $F_{n}(y)$ are plotted in Fig.~\ref{fig:Fn} (see also
Fig.~2 of \cite{GRW}). Their most salient features are as follows.
At moderate $y\lesssim1$, we have $F_n(y)=\mathcal{O}(1)$,
$\footnote{For $n=2$ there is mild logarithmic growth.}$ the
integral (\ref{eq:Fn}) receiving contributions from $x\sim1$. On the
other hand, for $y\gg1$ the integral has a saddle point at
$x_\ast=(n/y^n)^\frac
{1}{n+1}\ll1$, and the amplitude decays:%
\[
|F_{n}(y)|\simeq\frac{n^{\frac{1}{n+1}}}{\sqrt{n+1}}\,y^{-\frac{n+2}{n+1}%
}\quad(y\gg1)\,.
\]

The appearance of the scale $b_{c}$ is a peculiar feature of
T-scattering for $n>0$. Since the amplitude is the largest in the
region $y\lesssim1$, a typical scattering will have
$|\mathbf{q}|\lesssim b_{c}^{-1}$. Yet a classical particle
trajectory for these $\mathbf{q}$ is undefined, all impact
parameters $b\sim b_{c}$ contributing to the scattering. On the
other hand, for $b_{c}|\mathbf{q}|\gg1$ $(y\gg1)$ the scattering is
dominated by a characteristic impact parameter
$b_{\ast}=b_{c}x_{\ast}$, corresponding to the above saddle point.
In this case the particle trajectory is well defined and the
T-scattering is truly semiclassical, with many gravitons being
exchanged.

\subsection{'t Hooft's method}

\label{sec:tHooft}

An alternative computation of the small angle T-scattering amplitude can be
given using a method due to 't Hooft \cite{tHooft}, originally formulated in
four dimensions. In this approach, particle $B$ scatters on the classical
gravitational field created by particle $A$. In other words, particle $A$ is
treated as a classical point particle, while particle $B$ is treated quantum-mechanically.

Consider then the gravitational field of a relativistic classical point
particle $A$ of energy $E_{A}$ propagating in the positive $z$ direction. This
field is the $D$-dimensional generalization of the AS \cite{AS} shock wave:%
\begin{equation}
ds^{2}=-dx^{+}dx^{-}+\Phi(x_{\perp})\delta(x^{-})(dx^{-})^{2}+dx_{\perp}%
^{2}\,. \label{eq:AS}%
\end{equation}
Here $x^{\pm}=t\pm z,$ while $x_{\perp}$ denotes $D-2$ transverse directions.
Einstein's equations with the lightlike source%
\[
T_{--}=E_{A}\,\delta(x^{-})\delta^{(D-2)}(x_{\perp})
\]
reduce to one linear equation for the shock wave profile $\Phi$:%
\begin{equation}
-\partial_{\perp}^{2}\Phi=16\pi G_{D}E_{A}\,\delta^{(D-2)}(x_{\perp})\,.
\label{eq:lapl}%
\end{equation}
The solution of this equation coincides with the eikonal phase
(\ref{eq:eikR}) per unit of particle $B$ energy:
\begin{equation}
\Phi(x_{\perp})=E_{B}^{-1}\chi(x_{\perp})\,\text{.} \label{eq:chiphi}%
\end{equation}

The right-moving particle $B$ is confined to the SM 3-brane, and its
wavefunction solves the Klein-Gordon equation in the metric induced on the
brane by the shock wave (\ref{eq:AS}). At $x^{-}<0$ the wavefunction is a
standard plane wave
\[
\phi(x)=\exp(ip_{B}.x)=\exp(-iE_{B}x^{+})\,.
\]

\begin{figure}[h]
\begin{center}
\includegraphics[
height=1.5in ]{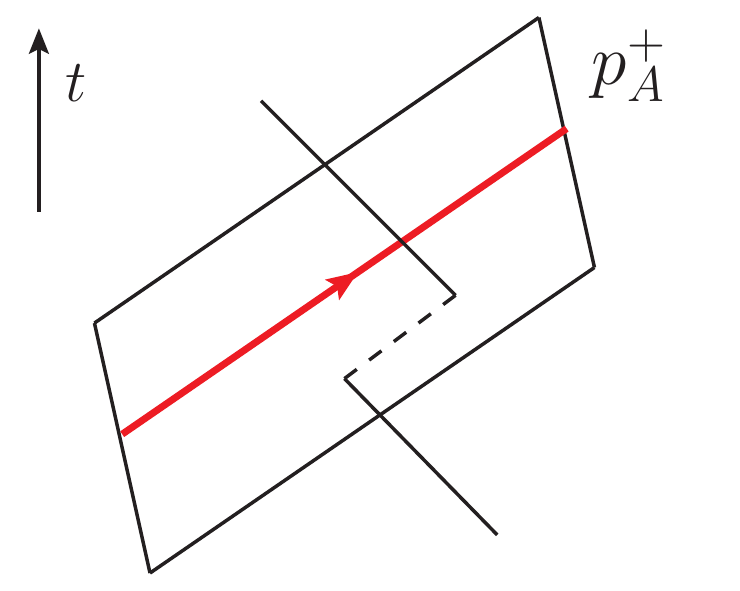}
\end{center}
\caption{\textit{The gravitational field of particle A (the AS shock wave) is
concentrated on the null plane $x^{-}=0$. Geodesics crossing this field
experience an $\mathbf{x}$-dependent shift of the $x^{+}$ coordinate. This is
the same shift as in Eq.\ (\ref{eq:after}).}}%
\label{fig:AS}%
\end{figure}

The metric (\ref{eq:AS}) has a strong discontinuity at $x^{-}=0$. To solve the
Klein-Gordon equation across the discontinuity, it is convenient to make a
coordinate transformation \cite{EG},\cite{R}%
\begin{align}
x^{-} & =\widetilde{x}^{-}\,,\nonumber\\
x^{+} & =\widetilde{x}^{+}+\theta(x^{-})\Phi(\widetilde{x}_{i})+x^{-}%
\theta(x^{-})\frac{(\partial\Phi(\widetilde{x}_{i}))^{2}}{4}\,,\label{eq:xt}\\
x_{i} & =\widetilde{x}_{i}+\frac{x^{-}}{2}\theta(x^{-})\partial_{i}%
\Phi(\widetilde{x}_{i})\,.\nonumber
\end{align}
In the $\widetilde{x}$ coordinates the metric is continuous across $x^{-}=0$.
When crossing the shock wave, the wavefunction remains continuous in these
coordinates. This means that for small positive $x^{-}$ we have:%
\begin{equation}
\phi(\widetilde{x})=\exp(-iE_{B}\widetilde{x}^{+})=\exp[-iE_{B}(x^{+}%
-\Phi(\mathbf{x})]\,. \label{eq:after}%
\end{equation}
The $\mathbf{x}$-dependent shift of the $x^{+}$ coordinate has a well-known
classical origin: it is related to the time delay experienced by geodesics
crossing the AS shock wave, see Fig.~\ref{fig:AS}.\footnote{We would like to
stress the \textit{auxiliary} character of the $\widetilde{x}$ coordinates, in
which the metric is not manifestly flat for $\widetilde{x}^{-}>0$, nor even
manifestly asymptotically flat, which makes these coordinates unsuitable for
defining asymptotic outgoing states. The asymptotic states should be described
in the $x$ coordinates, that's why in the last equation in (\ref{eq:after}) we
reverted to them.}

We now see from (\ref{eq:after}) that the wavefunction immediately before and
after the collision is related by a pure phase factor $\exp(iE_{B}%
\Phi(\mathbf{x})$), which via (\ref{eq:chiphi}) is identical with the eikonal
phase factor in (\ref{eq:eikR}). An alternative derivation, by directly
solving the Klein-Gordon equation, is given in Appendix \ref{sec:gluonAS}.

Thus, 't Hooft's method is equivalent to the resummation. This is not
surprising, because the external field approximation in quantum field theory
resums precisely crossed ladder diagrams \cite{WeinbergI}. The AS shock wave
is a solution to both linearized gravity and the full nonlinear Einstein's
equations. In retrospect, this explains why the diagrams in which gravitons
emitted by particles $A$ and $B$ interact did not have to be taken into
account in the resummation method. See \cite{Kabat:1992tb} for a detailed
discussion and comparison of the two methods in $D=4$.

Still, an attentive reader will notice two small differences between the two
results. First, Eq.~(\ref{eq:eikR}) contains $-1$ under the integral sign,
while 't Hooft's method gives a pure phase. This is the usual difference
between the S- and T-matrices, $S=1+iT$. Second, the amplitude (\ref{eq:eikR})
is relativistically normalized, while in the new derivation normalization
needs yet to be determined.

Modulo the normalization issue (which will be resolved in Section
\ref{sec:scalar} below), the power of 't Hooft's method relative to the
resummation is quite evident. The eikonal phase is given a simple physical
interpretation---it is related to the time delay experienced by geodesics upon
crossing the shock wave. The exponential factor $e^{i\chi}$ emerges as a whole
rather than by summing infinitely many individually large terms.

\section{T-scattering with hadrons: intuition and questions}

\label{sec:hadronic}

If TeV-scale gravity is the way of Nature, then transplanckian collisions may
be within the energy reach of the LHC. Moreover, transplanckian collisions may
be constantly happening in the atmosphere, between the atmospheric nucleons
and high-energy cosmic rays ($\sqrt{s}\sim10^{6}$ GeV for $E_{cr}\sim10^{11}$
GeV of the order of the GZK cutoff). In case of cosmic ray neutrinos this
signal could actually be observable.

Since protons are not elementary particles, the theory of small angle
T-scattering from Section \ref{sec:review} should be applied instead to
$2\rightarrow2$ collisions between the partonic constituents. Notice that
since we are dealing with CM energies well over a TeV, the typical momentum
transfers will be hard compared to the QCD scale, even though the scattering
angle has to be small for the eikonal approximation to be valid. Thus the
collision resolves the internal structure of the proton(s), and the partonic
picture is applicable \cite{Emparan:2001kf}.

Viewed another way, when two protons collide, there is a phase factor
$\sim\Phi(\mathbf{x}-\mathbf{y})$ for each pair of partons moving in the
opposite directions, see Fig.~\ref{fig:pp}. This factor tends to zero rapidly
at transverse separations $|\mathbf{x}-\mathbf{y}|\gg b_{c}$, where $b_{c}%
\sim(100$ GeV)$^{-1}$ for T-scattering at the LHC energies. Since partons are
distributed in the disk of radius $($GeV$)^{-1}\gg b_{c}$, it is unlikely that
more than one pair will undergo a hard collision.

We would like to briefly mention which observables one usually computes in
phenomenological studies. In $\nu p$ collisions one is mostly interested in
the total interaction cross section as a function of the energy transfer to
the proton \cite{Emparan:2001kf},\cite{Illana},\cite{Illana1}. We will discuss
a similar observable in Section \ref{sec:QCDeffects} below. On the other hand,
in the $pp$ collisions at the LHC one studies two jet final states of high
invariant mass, produced at a small angle to the beam \cite{GRW}%
,\cite{Lonnblad}, see Fig.~\ref{fig:DISpp}. These jets originate from all
possible parton pairs $(qq,qg,gg$) with the same partonic cross section, the
eikonal amplitude being independent of the particle spin. For $M_{D}$ not much
above a TeV, the dijet T-scattering signal turns out to be visible over the
QCD background.

So far it may look that from the point of view of QCD, the T-scattering is
just like any other hard process. Let us however discuss which parton
distribution factorization scale $\mu_{F}$ one should use when evaluating the
T-scattering cross sections---a necessary prerequisite for any practical computation.

For the usual hard processes, we are accustomed to the choice $\mu_{F}%
\sim|\mathbf{q|}$, but for the T-scattering this turns out to be
more subtle. As we discussed in Section \ref{sec:resum},
T-scattering becomes semiclassical in the region of large momentum
transfers $|\mathbf{q|}\gg b_{c}^{-1}$. In this regime, the
transverse distance characterizing the process is the typical impact
parameter $b_{\ast}\sim b_{c}/(b_{c}|\mathbf{q|})^{\frac{1}{n+1}}$
which is \textit{parametrically larger} than $|\mathbf{q}|^{-1}$. It
is for this reason that Ref.~\cite{Emparan:2001kf} advocated a
hybrid prescription: one should use $\mu_{F}\sim|\mathbf{q|}$ for
$|\mathbf{q|}\lesssim b_{c}^{-1}$ and switch to $\mu_{F}\sim
b_{\ast}^{-1}$ for $|\mathbf{q|}\gg b_{c}^{-1},$ see Eq.
(\ref{eq:mu}).\footnote{A limiting case of this prescription in the
context of black hole production was advocated earlier in the first
Ref.\ \cite{factory}.}

For T-scattering at the LHC energies, the factorization scale
$\mu_{F}(q)$ will be hard with respect to the QCD scale as long as
the momentum transfer $q$ is hard. This gives a self-consistency
check on the proposed picture.\footnote{Notice however that
$\mu_{F}(q)$ is a \textit{decreasing} function of the CM energy,
since the typical impact parameter grows with $\sqrt{s}$. Even for
hard momentum transfers, for sufficiently high $\sqrt{s}$ the
factorization scale will come down to a GeV, signalling a breakdown
of the partonic picture. At even higher CM energies (which are well
beyond the range of LHC or even cosmic ray collisions), the proton
should interact gravitationally as a point particle.}

The above is a summary of the current understanding of QCD effects in
T-scattering. Clearly, it is based mostly on intuition. We would like to
develop a systematic theory of these phenomena. In particular, such a theory
should allow to check the factorization scale proposal by a concrete
computation. We have to evaluate the leading log corrections to the
T-scattering cross section due to the initial state radiation emission, and to
show that they can be absorbed into a shift of the PDF factorization scale.
Since $\mu_{F}$ is conjectured to have a nontrivial dependence on $q$, some
nontrivial physics is likely to come out.

Two equivalent methods were given in Section \ref{sec:review} to describe
T-scattering without radiation. Which one shall we try to generalize to the
case when radiation is present?

For the resummation method, generalization does not seem to be easy, not even
for the one-gluon emission. Think about infinitely many crossed-ladder
diagrams, infinitely many places to attach the gluon line, and the necessity
to take into account the gravitational exchanges of the emitted gluon!

For 't Hooft's method, on the other hand, the situation looks hopeful: if only
particle $B$ radiates, it is quite clear how to include its radiation. Namely,
we should keep working in the classical gravitational background created by
particle $A$, but switch from relativistic quantum mechanics (wavefunctions,
the Klein-Gordon equation) to quantum field theory (Green's functions and
interaction vertices). We will follow this path and will see that it allows
relatively straightforward computations of the gluon emission amplitudes.

Physically, the assumption that particle $A$ does not QCD-radiate is realized
if $A$ is a lepton. If both $A$ and $B$ are strongly interacting, one could
first compute the radiation off $B$ (taking $A$ classical), then off $A$
(taking $B$ classical). Such an approximation of independent emission is valid
for the dominant, collinear, radiation in the usual perturbative processes.
For the T-scattering, we will be able to partially justify it below. But first
we have to understand well the case of non-radiating $A$.

\section{Quantum field theory in the shock wave background}

\label{sec:QFT}

\subsection{Scalar field}

\label{sec:scalar}

To compute the QCD radiation accompanying a transplanckian collision, we will
replace particle $A$ with the classical background it generates, but will keep
particle $B$ and the gluons as quantum fields. Thus we will be doing
perturbative QFT computations in the shock wave background. We start with the
simplest interacting QFT, the massless $\phi^{3}$ theory:%
\begin{equation}
\mathcal{L}=\sqrt{g}(\frac{1}{2}g^{\mu\nu}\partial_{\mu}\phi\partial_{\nu}%
\phi-\frac{\lambda}{3!}\phi^{3})\,. \label{eq:Lphi}%
\end{equation}
We will describe how to compute transition amplitudes in this theory, and how
these are related to the amplitudes in the full theory (i.e.~before particle
$A$ was replaced by a classical gravitational field).

The $g_{\mu\nu}$ in (\ref{eq:Lphi}) is the 4-dimensional metric obtained by
restricting the $D$-dimensional AS shock wave (\ref{eq:AS}) to the SM brane on
which both particles and the radiation propagate. We will continue using the
coordinates as in (\ref{eq:AS}), only restricting the number of $x_{\perp}$
components from $D-2$ to $2$. Two features make this theory much simpler than
it would be for generic curved backgrounds treated in \cite{BD}:

\begin{enumerate}
\item the metric is invariant under $x^{+}$ shifts. The conjugate momentum
$p^{-}$ is conserved. This leads in particular to the absence of spontaneous
particle creation.

\item the spacetime is flat except on the $x^{-}=0$ plane. The Feynman rules
are simplified by using the flat-space coordinates.
\end{enumerate}

We start by canonically quantizing the quadratic part of the lagrangian. The
scalar field modes are found by solving the equations of motion (EOM)\ in the
shock wave background with the plane wave conditions in the asymptotic past
$x^{-}<0$:\footnote{By boldface letters $\mathbf{p},\mathbf{x}$ we denote the
2-dimensional, transverse to the beam, part of 4-dimensional Lorentz vectors.
The Minkowski space signature is $-++\,+$.}%
\begin{align}
\phi_{p^{-},\mathbf{p}}^{\text{in}}(x) & =\theta(-x^{-})e^{i[p].x}%
+\theta(x^{-})\int\frac{d^{2}\mathbf{q}}{(2\pi)^{2}}\,I(p^{-},\mathbf{q}%
)\,e^{i[p+\mathbf{q}].x}\,,\label{eq:modes}\\
I(p^{-},\mathbf{q}) & \equiv\int d^{2}\mathbf{x}\,e^{-i\mathbf{q}%
.\mathbf{x}}e^{i\frac{1}{2}p^{-}\Phi(\mathbf{x})}\,.\nonumber
\end{align}
The compact \textquotedblleft vector in square brackets\textquotedblright%
\ notation denotes an on-shell 4-vector whose $+$ component is computed in
terms of the known $-$ and $\mathbf{\perp}$, i.e. $[p+\mathbf{q}%
]\equiv((\mathbf{p+q})^{2}/p^{-},p^{-},\mathbf{p+q}),$ etc.

The function $I(p^{-},\mathbf{q})$ is identical to the eikonal amplitude
(\ref{eq:eikR}), up to the normalization and the absence of $-1$ under the
integral sign (which means that it contains an extra $\delta$-function piece).

The modes (\ref{eq:modes}) solve the Klein-Gordon equation both for $x^{-}<0$
and for $x^{-}>0.$ Across the shock wave, they satisfy the matching condition
of Section \ref{sec:tHooft}:
\[
\phi(x^{-}=+\varepsilon,x^{+},\mathbf{x})=\phi(x^{-}=-\varepsilon,x^{+}%
-\Phi(\mathbf{x}),\mathbf{x})\,.
\]

We proceed to quantize the field by expanding in oscillators:%
\begin{gather*}
\phi(x)=\int_{p^{-}>0}\frac{dp^{-}d^{2}\mathbf{p}}{\sqrt{2p^{-}}(2\pi)^{3}%
}\left\{ a_{p^{-}\mathbf{p}}\phi_{p^{-}\mathbf{p}}^{\text{in}}(x)+a_{p^{-}%
\mathbf{p}}^{\dagger}[\phi_{p^{-}\mathbf{p}}^{\text{in}}(x)]^{\ast}\right\}
\,,\\[5pt]
\lbrack a_{p_{1}^{-}\mathbf{p}_{1}},a_{p_{2}^{-}\mathbf{p}_{2}}^{\dagger
}]=(2\pi)^{3}\delta(p_{1}^{-}-p_{2}^{-})\delta^{(2)}(\mathbf{p}_{1}%
-\mathbf{p}_{2})\,.
\end{gather*}
Such normalization of the creation/annihilation operators is standard for
quantizing on the light cone; it differs from the usual one by a simple rescaling.

Equivalently, we can quantize using\ the outgoing modes, which reduce to plane
waves for $x^{-}>0$:%
\begin{equation}
\phi_{p^{-},\mathbf{p}}^{\text{out}}(x)=\theta(x^{-})e^{i[p].x}+\theta
(-x^{-})\int\frac{d^{2}\mathbf{q}}{(2\pi)^{2}}\,I(p^{-},\mathbf{q}%
)\,e^{i[p-\mathbf{q}].x}\,. \label{eq:modes1}%
\end{equation}
The in and out modes are related by a unitary Bogoliubov transformation, which
acts only on the transverse momentum $\mathbf{p}$ but not on $p^{-}$. Thus
there is no spontaneous particle creation in this background; the vacuum is
unambiguously defined.

Let us now build a perturbation theory for transition amplitudes. The logic is
simplest in the position space. Even though the metric is singular at
~$x^{-}=0,$ it is easy to see that$\sqrt{g}\equiv1$: the metric determinant
drops out of the interaction lagrangian. Thus the Feynman diagrams will be
given by flat space integrals, with no singular contribution from the shock
wave. For instance, the $t$-channel diagram contributing to the $p_{1}%
,p_{2}\rightarrow p_{3},p_{4}$ transition amplitude will be given by:%
\begin{equation}
\raisebox{-0.3491in}{\includegraphics[
natheight=1.399700in,
natwidth=2.187900in,
height=0.8988in,
width=1.3937in
]{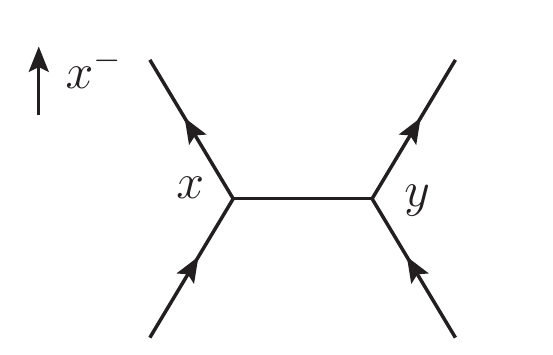}}=(-i\lambda)^{2}\int d^{4}x\,d^{4}y\,[\phi_{3}^{\text{out}%
}(x)]^{\ast}\,[\phi_{4}^{\text{out}}(y)]^{\ast}\,G(x,y)\,\phi_{1}^{\text{in}%
}(x)\,\phi_{2}^{\text{in}}(y)\,. \label{eq:exF}%
\end{equation}
The $\phi^{\text{out }}$ and $\phi^{\text{in}}$ enter as the in and out state
wavefunctions. The propagator $G(x,y)$ must be $x^{-}$-ordered:
\[
G(x,y)=\theta(x^{-}-y^{-})\left\langle 0|\phi(x)\phi(y)|0\right\rangle
+(x\leftrightarrow y)\,.
\]
In (\ref{eq:exF}) we have to integrate in all possible $x^{-}$ orderings of
$x$ and $y$ with respect to each other and to the shock wave sitting at
$x^{-}=0$. The propagator will take different forms depending on the ordering.
For $x^{-}$ and $y^{-}$ on the same side of the shock, we get the flat space
result:%
\[
G_{\text{flat}}(x,y)=\theta(x^{-}>y^{-}>0)\int_{p^{-}>0}\frac{dp^{-}%
d^{2}\mathbf{p}}{2p^{-}(2\pi)^{3}}e^{i[p].(x-y)}+(x\leftrightarrow y)\,.
\]
On the other hand, across the shock wave we have%
\[
G_{\text{cross}}(x,y)=\theta(x^{-}>0>y^{-})\int_{p^{-}>0}\frac{dp^{-}%
d^{2}\mathbf{p}}{2p^{-}(2\pi)^{3}}\int\frac{d^{2}\mathbf{q}}{(2\pi)^{2}%
}\,I(p^{-},\mathbf{q})\,e^{i[p+\mathbf{q}].x-i[p].y}+(x\leftrightarrow y)\,.
\]

The momentum-space Feynman rule can now be found by straightforward Fourier
transformation; they are as follows. The in and out states are specified by
the $p^{-}$ and $\mathbf{p}$ of all incoming and outgoing particles. The
transition amplitude $i\rightarrow f$ in the external gravitational field of
particle $A$ is then given by:%
\[
_{\text{out}}\left\langle f|i\right\rangle _{\text{in}}=2(2\pi)\,\delta
(p_{f}^{-}-p_{i}^{-})\mathcal{M}(i\rightarrow f)\,.
\]
The $\mathcal{M}(i\rightarrow f)$ is a function of the external momenta
computed as a series in $\lambda$ according to the following rules. To obtain
the $O(\lambda^{N})$ term:

\begin{itemize}
\item Draw the Feynman diagrams with $N$ $\phi^{3}$ vertices, considering all
possible $x^{-}$-orderings of these vertices with respect to each other and to
the shock wave at $x^{-}=0$.

\item Consider all shock wave crossings as additional vertices, with entering
transverse momenta $\mathbf{q}_{a}$ representing momentum exchange with the
shock wave.

\item Assign $p^{-}$,$\mathbf{p}$ internal lines momenta by using their
conservation in all vertices ($\phi^{3}$ and shock wave crossings). Momentum
flow is in the direction of increasing $x^{-}$. The internal $p^{+}$ momenta
are not conserved but are assigned by using the on-shell condition
$p^{+}=\mathbf{p}^{2}/p^{-}.$

\item For each $\phi^{3}$ vertex multiply by $-i\lambda.$

\item For each shock wave crossing vertex multiply by $p^{-}I(p^{-}%
,\mathbf{q}_{a})$, where $p^{-}$ is conserved in the crossing.

\item For each internal line (i.e.~a line connecting two vertices, $\phi^{3}$
or shock wave crossing) carrying momentum $p^{-}$, multiply by $\theta
(p^{-})/p^{-}.$

\item The $\phi^{3}$ vertices and the shock wave at $x^{-}=0$ divide the
$x^{-}$ axis into two unbounded and $N$ bounded intervals. For each bounded
interval, we define an\textit{ intermediate state}, consisting of all the
particles whose internal lines traverse this interval. For each intermediate
state at \textit{negative} $x^{-}$, the amplitude is multiplied by%
\[
\frac{i}{\sum_{i}p^{+}-\sum_{\text{interm}}p^{+}+i\varepsilon}\,.
\]
For each intermediate state at \textit{positive} $x^{-}$, it is multiplied by%
\[
\frac{i}{\sum_{f}p^{+}-\sum_{\text{interm}}p^{+}+i\varepsilon}\,.
\]
The sums are over all particles in the initial ($x^{-}=-\infty$),
intermediate, and final ($x^{-}=+\infty$) state.

\item Integrate over the momenta $\mathbf{q}_{a}$ exchanged with the shock
wave:%
\[
\int(2\pi)^{2}\delta^{(2)}(\sum\mathbf{q}_{a}+\mathbf{p}_{i}-\mathbf{p}%
_{f})\prod\frac{d^{2}\mathbf{q}_{a}}{(2\pi)^{2}}\,.
\]

\item For loop diagrams, integrate over all undetermined momenta
$(k^{-},\mathbf{k})$:%
\[
\int\frac{dk^{-}d^{2}\mathbf{k}}{2(2\pi)^{3}}\,.
\]

\end{itemize}

The reader will notice a striking similarity to the usual light-cone
perturbation theory (PT) rules \cite{LC}. Notice in particular the light-cone
energy denominators, and the $\theta(p^{-})$ factors, which eliminate some of
the diagrams present in the time-ordered `old' perturbation theory. New
features in our case are the shock wave crossing vertices, and that there are
two types of energy denominators, depending on the ordering with respect to
the shock wave. We thus have `light-cone PT in presence of an instantaneous
interaction'. Many years ago, Bjorken, Kogut and Soper \cite{BKS} have
developed light-cone PT in external electromagnetic field, and argued that at
sufficiently high energies interaction with the external field can be
represented as an instantaneous eikonal scattering.\footnote{We thank Zoltan
Kunszt for bringing this work to our attention.} In our case, the eikonal
factor has gravitational origin, but the formalism is the same. The formalism
of \cite{BKS} has found application in the dipole scattering approach to the
DIS at small $x$: an almost-real photon splits into two quarks which then
undergo eikonal scattering in the gluon field of the proton \cite{hebecker}.
The difference is that the gluon field of the proton is not really known,
while in our case the eikonal phase can be computed exactly.

We will now demonstrate the rules by computing a couple of amplitudes. The
elastic one-particle amplitude $\mathcal{M}(p\rightarrow p^{\prime})$ is given
by just one diagram with a shock wave crossing vertex (denoted by a cross):%
\begin{gather}
\includegraphics[
natheight=1.736800in,
natwidth=1.484600in,
height=1.1158in,
width=0.9563in
]{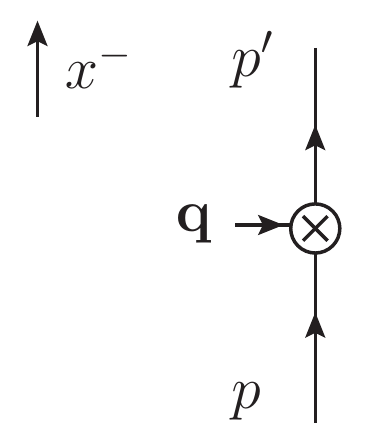}\nonumber\\
\mathcal{M}(p\rightarrow p^{\prime})=p^{-}I(p^{-},\mathbf{p}-\mathbf{p}%
^{\prime})\qquad(p^{2}=p^{\prime2}=0,\,p^{-}=p^{\prime-})~. \label{eq:MAel}%
\end{gather}

As a more complicated example, let us compute one of the diagrams appearing in
the computation of the amplitude $p_{1},p_{2}\rightarrow p_{3},p_{4}%
$:\footnote{In physical applications below we will be computing amplitudes
with one incoming and several outgoing particles. This amplitude containing
two incoming particles is considered for illustrative purposes only.}%
\[
\includegraphics[
natheight=2.828600in,
natwidth=4.571400in,
height=1.9195in,
width=3.091in
]{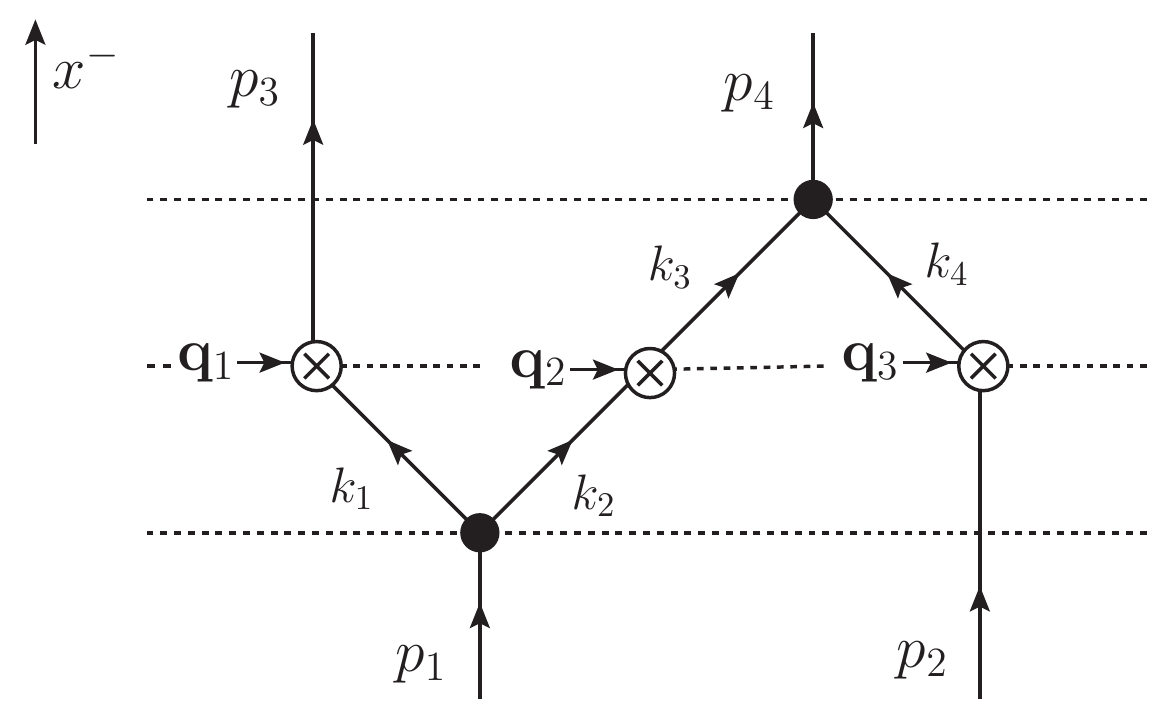}
\]
We have two $\phi^{3}$ vertices and three shock wave crossings. The dotted
lines stress the $x^{-}$ ordering of the vertices. The $k_{i}$ are the
internal line momenta, whose $-$ and $\perp$ components are fixed via momentum
conservation, while the $+$ components are determined by being on shell. There
are two intermediate states: one before the shock wave $(k_{1},k_{2},p_{2})$,
and one after $(p_{3},k_{3},k_{4})$. The value of this diagram is thus:%
\begin{align*}
& \int(2\pi)^{2}\delta^{(2)}(\sum\mathbf{q}_{a}+\mathbf{p}_{1}+\mathbf{p}%
_{2}-\mathbf{p}_{3}-\mathbf{p}_{4})\prod_{a=1}^{3}\frac{d^{2}\mathbf{q}_{a}%
}{(2\pi)^{2}}\,\\
& \times(-i\lambda)^{2}\,k_{1}^{-}I(k_{1}^{-},\mathbf{q}_{1})\,k_{2}%
^{-}I(k_{2}^{-},\mathbf{q}_{2})\,p_{2}^{-}I(p_{2}^{-},\mathbf{q}_{1})\,\\
& \times\frac{i}{(p_{1}^{+}+p_{2}^{+})-(k_{1}^{+}+k_{2}^{+}+p_{2}%
^{+})+i\varepsilon}\frac{i}{(p_{3}^{+}+p_{4}^{+})-(p_{3}^{+}+k_{3}^{+}%
+k_{4}^{+})+i\varepsilon}\prod_{i=1}^{4}\frac{\theta(k_{i}^{-})}{k_{i}^{-}}\,.
\end{align*}

Finally, we have to discuss the relation between the transition amplitude
$\mathcal{M}(i\rightarrow f)$ and the full relativistic scattering amplitude
$\mathcal{M}_{\text{rel}}$, i.e.~the one obtained when we reinstate particle
$A$ as a quantum particle as opposed to replacing it with its classical field.
We have:%
\begin{equation}
\mathcal{M}_{\text{rel}}(A+i\rightarrow A^{\prime}+f)=-i\,2p_{A}%
^{+}\,\mathcal{M}(i\rightarrow f)\,.\, \label{eq:MMA}%
\end{equation}
The incoming (outgoing) momenta of particle $A$ are assigned as follows:%
\[
p_{A}=(p_{A}^{+},0,0),\quad p_{A}^{\prime}=(p_{A}^{+},0,\mathbf{p}%
_{i}-\mathbf{p}_{f})\,.
\]
In other words, particle $A$ absorbs the total transverse momentum exchanged
with the shock wave. As long as the momentum transfer is small compared to
$p_{A}^{+}$, $A^{\prime}$ is almost on shell and the approximation is justified.

The relative factor $-i\,2p_{A}^{+}\,$ in (\ref{eq:MMA}) is related to the
normalization of the particle $A$ state, which is lost when we replace it with
a classical field. This factor is thus process-independent. For instance one
can extract it from the external field approximation in QED \cite{WeinbergI}.
The extra $i$ can be traced back to the external field creation vertex, which
carries a factor of $i$.

This settles the question of relativistic normalization of the amplitudes
computed via 't Hooft's method. We can now complete the comparison with the
resummation method. Using Eqs.~(\ref{eq:MAel}), (\ref{eq:MMA}) we have%
\begin{equation}
\mathcal{M}_{\text{rel}}(A+B\rightarrow A^{\prime}+B^{\prime})=-2ip_{A}%
^{+}p_{B}^{-}\,I(p_{B}^{-},\mathbf{q})\,, \label{eq:nog}%
\end{equation}
which agrees with the eikonal amplitude from Eq.~(\ref{eq:AeikR}) including
the normalization, modulo the difference between the S- and T-matrices already
discussed in Section \ref{sec:tHooft}.

\subsection{Gauge field}

\label{sec:gauge}

In order to keep technical details to a minimum, we will not consider
fermionic fields in the shock wave background. Instead, we will stick to a toy
model in which charged matter (partonic constituents of the colliding hadrons)
consists of massless scalars. This will be sufficient given our general goals.
On the other hand, since the coupling constant of the $\phi^{3}$ lagrangian
has dimension of mass, the cubic self-interaction is not a good model for the
QCD radiation. We do have to introduce gauge fields. Thus we switch from
(\ref{eq:Lphi}) to a different microscopic lagrangian, describing the SU(3)
Yang-Mills theory and a massless complex scalar in the fundamental
representation, propagating in the shock wave background:%
\begin{equation}
\mathcal{L}=\sqrt{g}(\frac{1}{2}g^{\mu\nu}(D_{\mu}\phi)^{\ast}D_{\nu}%
\phi-\frac{1}{2}\text{Tr}[F^{\mu\nu}F_{\lambda\sigma}])\,. \label{eq:Lsg}%
\end{equation}

Most of the $\phi^{3}$ formalism is carried over with trivial changes. Instead
of repeating the whole discussion, we will introduce the necessary
modifications on a concrete example.

Namely, let us consider the one-gluon emission: particle $B$, while scattering
in the gravitational field of particle $A$, emits a gluon. The amplitude is
given by the sum of the following two diagrams:%
\begin{equation}
\text{(I)}{\includegraphics[
natheight=2.171600in,
natwidth=2.380900in,
height=1.4958in,
width=1.6373in
]{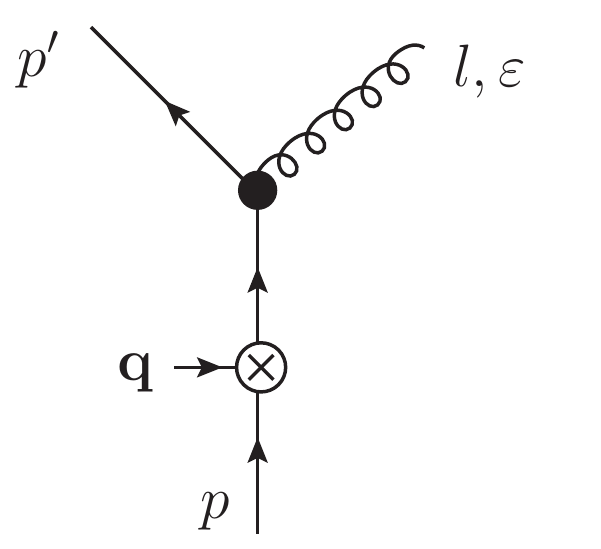}}\quad(\text{II)}{\includegraphics[
natheight=2.171600in,
natwidth=2.987300in,
height=1.4563in,
width=1.9975in
]{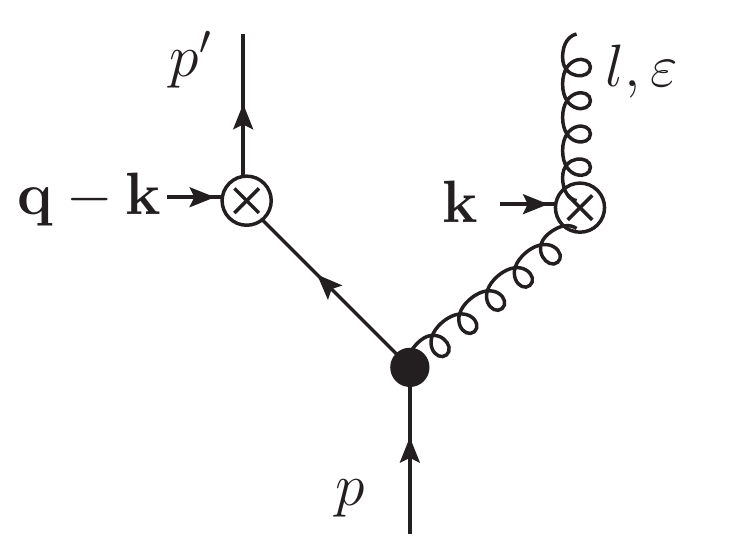}} \label{eq:1g}%
\end{equation}
The new objects are the gluon emission and the gluon shock wave crossing vertices.

To simplify the computations, we will impose the Lorenz and light-cone gauge
conditions:%
\[
D^{\mu}A_{\mu}=0,\quad A_{+}=0\,.
\]
The treatment in a general gauge and demonstration of gauge invariance is
given in Appendix \ref{sec:gluonAS}.

Gluon emission in curved space is described by the cubic term in the
lagrangian:%
\begin{equation}
ig_{s}\int d^{4}x\sqrt{g}g^{\mu\nu}\phi_{i}^{\ast}\overleftrightarrow
{\partial}_{\mu}\phi_{j}(T^{a})_{ij}\,A_{\nu}^{a}\,. \label{eq:gluonEm}%
\end{equation}
Here $g_{s}$ is the strong coupling constant, and the SU(3) generators are
normalized by Tr$(T^{a}T^{b})=1/2$. In the light-cone gauge the singular
component $g^{++}\propto\delta(x^{-})$ drops out (see Appendix
\ref{sec:gluonAS} for a more detailed discussion). The gluon emission vertex
is thus the same as in flat space:%
\begin{gather}
\raisebox{-0.3988in}{\includegraphics[
natheight=1.403100in,
natwidth=1.432300in,
height=0.9683in,
width=0.988in
]{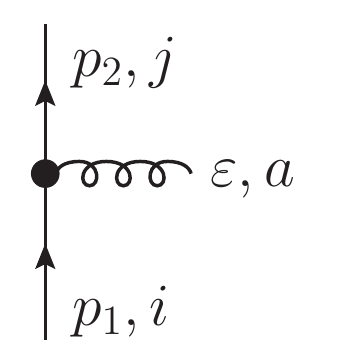}}\nonumber\\
g_{s}T_{ij}^{a}(p_{1}+p_{2}).\varepsilon\quad\longrightarrow\quad~g_{s}%
T_{ij}^{a}(\mathbf{p}_{1}+\mathbf{p}_{2}-\frac{p_{1}^{-}+p_{2}^{-}}{l^{-}%
}\mathbf{l}).\boldsymbol{\upepsilon\qquad(}\varepsilon_{+}=0,\,l.\varepsilon
=0)\,, \label{eq:gluonv}%
\end{gather}
where we used the Lorenz gauge to eliminate the $\varepsilon_{-}$ component.

The gluon shock wave crossing vertex contains the same factor $p^{-}%
I(p^{-},\mathbf{q})$ as in the scalar case. A new feature is that the
$\varepsilon_{-}$ polarization component changes in the crossing according to:%
\begin{equation}
\raisebox{-0.3988in}{\includegraphics[
natheight=1.567800in,
natwidth=1.833700in,
height=0.8465in,
width=0.988in
]{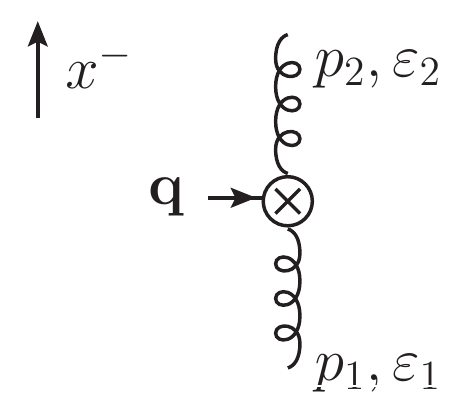}}\quad\varepsilon_{2-}=\varepsilon_{1-}-\frac
{\boldsymbol{\upepsilon}_{1}.\mathbf{q}}{p_{1}^{-}},~\boldsymbol{\upepsilon}%
_{2}=\boldsymbol{\upepsilon}_{1}\quad(\varepsilon_{+}\equiv0)\,,
\label{eq:gluoncross}%
\end{equation}
This rule is easy to guess from consistency with the imposed gauge;\ see
Appendix \ref{sec:gluonAS} for an explicit derivation. Notice however that we
don't have to keep track of this change in $\varepsilon_{-}$ if we use the
simplified gluon emission vertex in (\ref{eq:gluonv}).

We are now ready to evaluate the above two diagrams. Working for simplicity in
the frame where $\mathbf{p}=0$, we get:%
\begin{align}
\mathcal{M}^{(\text{I)}} & =ig_{s}T_{ij}^{a}\,I(p^{-},\mathbf{q}%
)\frac{(2\mathbf{p}^{\prime}+\mathbf{l}-\frac{2p^{\prime-}+l^{-}}{l^{-}%
}\mathbf{l}).\boldsymbol{\upepsilon}}{p^{\prime+}+l^{+}-[p^{\prime}+l]^{+}%
}\boldsymbol{\,},\nonumber\\
\mathcal{M}^{(\text{II})} & =ig_{s}T_{ij}^{a}\int\frac{d^{2}\mathbf{k}%
}{(2\pi)^{2}}I(l^{-},\mathbf{k})\,I(p^{\prime-},\mathbf{q}-\mathbf{k}%
)\frac{(\mathbf{k}-\mathbf{l}-\frac{2p^{-}-l^{-}}{l^{-}}(\mathbf{l}%
-\mathbf{k})).\boldsymbol{\upepsilon}}{-[p-l+\mathbf{k}]^{+}-[l-\mathbf{k}%
]^{+}+i\varepsilon}\boldsymbol{\,}. \label{eq:M1geval}%
\end{align}
Physical consequences of the derived expressions will be discussed below.

As a final comment, we note that lagrangian (\ref{eq:Lsg}) contains also a
cubic gluon self-interaction vertex, which could be discussed analogously to
(\ref{eq:gluonv}), as well as two quartic vertices ($\phi\phi AA$ and $AAAA$).
The quartic vertices do not contribute to the amplitudes in the collinearly
enhanced region, and we will not need their precise expressions.

\section{Initial state radiation in T-scattering}

\label{sec:QCDeffects}

The dominant QCD radiation effects in the usual perturbative hard scattering
processes are the collinear initial and final state radiation. We now proceed
to see how these effects manifest themselves in the T-scattering. We will
focus on the initial state radiation and its effect on the parton distribution
scale. Final state radiation, which happens after the partons cross the shock
wave, is expected to be as usual.

\subsection{Observable}

To discuss radiative corrections to the PDFs, we need to choose a process and
an observable which can be defined and computed beyond LO. The simplest such
process is the T-scattering analogue of the DIS. In other words, we will
consider an electron-proton T-scattering $ep\rightarrow e$+anything at a fixed
momentum transfer. This is like in Fig.~\ref{fig:DISpp} with an electron
instead of a neutrino.
%We will comment below on other observables, such as
%dijets in the $pp$ scattering.

The scattering is characterized by $t=-q^{2}$ and $q^{+}$, the energy transfer
to the proton. These can be measured by observing the electron. As usual, we
assume small angle scattering: $|t|\ll s$. We will also assume that the
relative electron energy loss is small, $q^{+}\ll p_{A}^{+}$. Under these
conditions, and also since the electron does not QCD-radiates, we can
represent it by a classical relativistic point particle of fixed energy. This
is our `particle $A$'. Using the $A^{\prime}$ on shell condition
$(p_{A}-q)^{2}=0$, it is easy to show that the momentum transfer is mostly in
the transverse plane, as expressed by the relation:%
\[
\mathbf{q}^{2}=q^{2}\left( 1+2q^{+}/p_{A}^{+}\right) \simeq q^{2}\,.
\]

Like in the DIS, we are interested in the differential cross section with
respect to $\mathbf{q}$ and the Bjorken $x$:
\[
\frac{d\sigma}{d^{2}\mathbf{q}\,dx},\quad x=\frac{q^{2}}{p_{B}^{-}q^{+}}%
,\quad0<x<1\,.
\]
As is customary, we will first analyze the partonic cross section $\hat
{\sigma}$ between the electron and a quark (particle $B$). We will work in a
toy model of scalar quarks. At LO (no gluon emission), the amplitude is
(\ref{eq:nog}) and the partonic cross section is given by
\[
\frac{d\hat{\sigma}_{\text{LO}}}{d^{2}\mathbf{q}\,dx}=\delta(x-1)\frac{1}%
{4\pi^{2}}|I(p_{B}^{-},\mathbf{q})|^{2}\,.
\]

\subsection{One gluon emission in momentum space}

Armed with the formalism from Section \ref{sec:QFT}, we can easily
write down the gluon emission amplitudes. At the NLO we have
diagrams with real gluon emission, as in Eq.~(\ref{eq:1g}), as well
as virtual corrections to the external legs and the vertices in the
elastic amplitude. As usual, the latter diagrams do not have to be
computed explicitly, since they only correct the coefficient of
$\delta(x-1).$\footnote{Vertex corrections play a role in cancelling
the IR divergence in the cross section corresponding to the
emissions of soft gluons at large angles. This cancellation holds
for any hard process. Specifically, it will also happen for the
T-scattering because soft gluons do not feel the shock wave of
particle A. Thus we do not discuss soft gluons in what follows,
concentrating on the collinear divergence.} We thus focus on the
real emission.

The partonic cross section with one gluon emitted is given by a phase space
integral (see Appendix \ref{sec:phase})%
\begin{gather}
\frac{d\hat{\sigma}_{\text{NLO}}}{d^{2}\mathbf{q}\,dx}=\frac{1}{16\pi^{2}%
\hat{s}}\int\frac{d^{2}\mathbf{l}}{2(2\pi)^{3}}\int_{0}^{1}\frac{dz}%
{z(1-z)}\,\delta(x-\mathbf{q}^{2}/(p_{B}^{-}q^{+}))\,|\mathcal{M}_{\text{rel}%
}|^{2},\label{eq:1gpart}\\[5pt]
q^{+}=\mathbf{l}^{2}/l^{-}+(\mathbf{q}-\mathbf{l})^{2}/p_{B^{\prime}}%
^{-}\,.\nonumber
\end{gather}
Here $\mathcal{M}_{\text{rel}}$ is the relativistic scattering amplitude
$A+B\rightarrow A^{\prime}+B^{\prime}+g$, related to the transition amplitude
in the external field $\mathcal{M}(B\rightarrow B^{\prime}+g)$ via Eq.
(\ref{eq:MMA}). The amplitude $\mathcal{M}$ is in turn the sum of the two
diagrams (\ref{eq:1g}), evaluated in Eq.~(\ref{eq:M1geval}).

We are using notation from (\ref{eq:1g}) with $p\equiv p_{B}$, $p^{\prime
}\equiv p_{B^{\prime}}$. The $q^{+}$ is the total + momentum of the
quark-gluon system after the collision. The $z$ is the $p^{-}$ momentum
fraction carried off by quark $B^{\prime}$:%
\[
p^{\prime-}=zp^{-},\quad l^{-}=(1-z)p^{-}\,.
\]

\begin{figure}[h]
\begin{center}
\includegraphics[
natheight=2.025800in, natwidth=1.747100in, height=2.0824in,
width=1.7994in ]{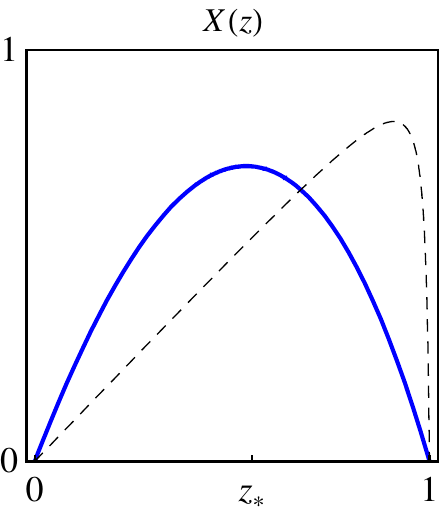}
\end{center}
\caption{\textit{The function }$X(z)$\textit{ for a generic }$\mathbf{l}%
$\textit{ (solid line) and for }$|\mathbf{l}|\ll|\mathbf{q}|$\textit{ (dashed
line).}}%
\label{fig:Xz}%
\end{figure}

Let us first analyze which region of the $\mathbf{l}$ plane contributes to the
integral (\ref{eq:1gpart}). For which $\mathbf{l}$ is there a $z$ saturating
the $\delta$-function? The relevant function (see Fig. \ref{fig:Xz})
\[
X(z)\equiv\mathbf{q}^{2}/(p_{B}^{-}q^{+})\equiv\frac{\mathbf{q}^{2}%
}{(\mathbf{q-l})^{2}/z+\mathbf{l}^{2}/(1-z)}\,,
\]
has a maximum value%
\[
\max_{0<z<1}X(z)=X(z_{\ast})=\frac{\mathbf{q}^{2}}{(|\mathbf{q-l}%
|+|\mathbf{l}|)^{2}}\quad\text{(}z_{\ast}=\frac{|\mathbf{q-l}|}{|\mathbf{q-l}%
|+|\mathbf{l}|})\,.
\]
Thus, the integrand of (\ref{eq:1gpart}) is nonzero for $\mathbf{l}$ belonging
to the ellipse:%
\[
|\mathbf{q-l}|+|\mathbf{l}|<|\mathbf{q}|/\sqrt{x}\,.
\]
In other words, phase space limits the transverse momentum of the emitted
gluon to be at most $\mathcal{O}(\mathbf{q})$.

Let us now examine the amplitude, whose two parts are given in
Eq.~(\ref{eq:M1geval}). By analogy with the usual DIS, we expect that part
(I), corresponding to the gluon emission \textit{after} the hard scattering,
gives only a finite correction to the cross section, while part (II) contains
a logarithmic IR divergence which has to be absorbed by redefining the PDFs.
Let us see formally how this happens.

Notice that part (I) of the amplitude is non-singular in the $\mathbf{l}$
plane. In particular, the intermediate state denominator is completely fixed
at $\mathbf{q}^{2}(\frac{1}{x}-1)/p^{-}>0$. Omitting the $x$ dependent
factors, the amplitude is thus $\mathcal{O}(g_{s}\,I\,\mathbf{q}%
.\boldsymbol{\upepsilon/}\mathbf{q}^{2}),$ and its square is $\mathcal{O}%
(g_{s}^{2}|I|^{2}/\mathbf{q}^{2}).$ After integrating over the ellipse in the
$\mathbf{l}$ plane (area $\sim\pi\mathbf{q}^{2}),$ we get a finite
contribution to the differential cross section of the relative order
$\mathcal{O}(\alpha_{s}/\pi).$ This is as expected.

Interesting physics is associated with part (II), whose expression can be
simplified as follows:%
\begin{align}
\mathcal{M}^{(\text{II})} & =-ig_{s}T_{ij}^{a}\,2p^{-}%
z\,\boldsymbol{\upepsilon}.\mathbf{M\,,}\nonumber\\[5pt]
\mathbf{M}_{i} & =\int\frac{d^{2}\mathbf{k}}{(2\pi)^{2}}\frac{(\mathbf{k}%
-\mathbf{l})_{i}}{(\mathbf{k}-\mathbf{l})^{2}}\,I(zp^{-},\mathbf{q}%
-\mathbf{k})\,I((1-z)p^{-},\mathbf{k})\,\label{eq:Mi0}\\
& \equiv\frac{-\mathbf{l}_{i}}{\mathbf{l}^{2}}\,\widetilde{I}(zp^{-}%
,\mathbf{q})+\,\frac{(\mathbf{q}-\mathbf{l})_{i}}{(\mathbf{q}-\mathbf{l})^{2}%
}\widetilde{I}((1-z)p^{-},\mathbf{q})\nonumber\\
& +\int\frac{d^{2}\mathbf{k}}{(2\pi)^{2}}\frac{(\mathbf{k}-\mathbf{l})_{i}%
}{(\mathbf{k}-\mathbf{l})^{2}}\,\widetilde{I}(zp^{-},\mathbf{q}-\mathbf{k}%
)\,\widetilde{I}((1-z)p^{-},\mathbf{k})\,. \label{eq:Mi}%
\end{align}
Here we separated the regular part $\widetilde{I}$ of the $I(p^{-}%
,\mathbf{q})$ from the $\delta$-function piece describing the propagation
without scattering:%
\begin{gather*}
I(p^{-},\mathbf{q})=(2\pi)^{2}\delta^{(2)}(\mathbf{q})+\widetilde{I}%
(p^{-},\mathbf{q})\,,\\[5pt]
\widetilde{I}(p^{-},\mathbf{q})=2\pi b_{c}^{2}F_{n}(b_{c}|\mathbf{q}|)\,,
\end{gather*}
where $F_{n}(y)$ are the same functions as in Eq.~(\ref{eq:Fn}). We omitted a
total $\delta^{(2)}(\mathbf{q})$ piece from (\ref{eq:Mi}).

The physical meaning of the decomposition (\ref{eq:Mi}) is as follows. In the
first two terms, only one of the two splitting products of quark $B$
participates in the gravitational interaction, the other one passing the shock
wave without scattering. The last term instead describes their coherent
gravitational scattering, as in Fig.~\ref{fig:DISg}. We call it the
\textit{rescattering} term, since it corresponds to the situation when the
emitted QCD radiation changes its direction in the field of the shock wave.

We now proceed to studying corrections to the cross section. Consider first
the case $|\mathbf{q}|\lesssim b_{c}^{-1}.$ In this case all the entering
$\widetilde{I}$ functions are $\mathcal{O}(2\pi b_{c}^{2}).$ The rescattering
term can be estimated by integrating up to $|\mathbf{k}|\sim b_{c}^{-1}$
beyond which point the $\widetilde{I}$ decrease faster than $|\mathbf{k}%
|^{-1}$, and the integral converges. We get%
\begin{equation}
|\mathbf{M}_{\text{resc}}|\sim\frac{\pi b_{c}^{-2}}{(2\pi)^{2}}b_{c}(2\pi
b_{c}^{2})^{2}\sim b_{c}(2\pi b_{c}^{2})<\frac{1}{|\mathbf{q}|}(2\pi b_{c}%
^{2})\qquad(|\mathbf{q}|\lesssim b_{c}^{-1})\,. \label{eq:Mresc1}%
\end{equation}
We see that rescattering is subleading to the first two terms in
(\ref{eq:Mi})$.$

Concentrating on the first two terms, the dominant contribution to the cross
section comes from the singularities at $\mathbf{l}\rightarrow0$ and
$\mathbf{l}\rightarrow\mathbf{q}$. Squaring the amplitude and integrating we
get:
\begin{gather}
\frac{d\hat{\sigma}_{\text{NLO}}}{d^{2}\mathbf{q}\,dx}\simeq\frac{1}{4\pi^{2}%
}\left\{ \bigl|\widetilde{I}(xp^{-},\mathbf{q})\bigr|^{2}P_{Q\rightarrow
Q}(x)+\bigl|\widetilde{I}(xp^{-},\mathbf{q})\bigr|^{2}P_{Q\rightarrow
g}(x)\right\} \frac{\alpha_{s}}{2\pi}\log\frac{\mathbf{q}^{2}}{\mu
_{\text{IR}}^{2}}\,\qquad(|\mathbf{q}|\lesssim b_{c}^{-1}\text{)},\nonumber\\
P_{Q\rightarrow Q}(x)=C_{F}\frac{2x}{1-x},\quad P_{Q\rightarrow g}%
(x)=P_{Q\rightarrow Q}(1-x)\,,\quad C_{F}=4/3. \label{eq:NLO}%
\end{gather}
Here we used that in the relevant regions of integration (see Fig.
\ref{fig:Xz})%
\[
x=X(z)\simeq z\quad(|\mathbf{l}|\ll|\mathbf{q}|)\,,\qquad x=X(z)\simeq
1-z\quad(|\mathbf{q}-\mathbf{l}|\ll|\mathbf{q}|)\,.
\]
Eq. (\ref{eq:NLO}) has the standard factorized form expected from an NLO QCD
correction to a hard scattering \cite{QCD}. The IR divergent logarithm
multiplies the quark-electron and gluon-electron LO cross sections, with the
\textit{scalar} quark DGLAP splitting functions $P_{Q\rightarrow Q}$ and
$P_{Q\rightarrow g}$ as coefficients\footnote{See \cite{Kiselev} for the
splitting functions of a colored scalar.}. As usual, we can absorb the IR
divergence into the quark (first term) and gluon (second term) PDFs. If we fix
the parton distribution scale at the upper cutoff, $\mu_{F}^{2}=\mathbf{q}%
^{2}$, then the whole logarithmic correction is absorbed.

It is of course not surprising that we managed to recover the standard
factorization for \thinspace$|\mathbf{q}|\lesssim b_{c}^{-1}$: rescattering
was not important in this case, and without rescattering there is no
difference between transplanckian and any other hard scattering.

Let us proceed to the case $|\mathbf{q}|\gg b_{c}^{-1}$. The situation here is
more complicated since the rescattering is no longer subleading. Consider for
example the region $|\mathbf{q}|\gg|\mathbf{l}|\gg b_{c}^{-1}$. The
rescattering integral is dominated by $|\mathbf{k}|\lesssim b_{c}^{-1}$, where
$\widetilde{I}((1-z)p^{-},\mathbf{k})$ is maximal, and not, say, by the region
of $|\mathbf{k}|\sim|\mathbf{l}|,$ where $\frac{(\mathbf{k}-\mathbf{l})_{i}%
}{(\mathbf{k}-\mathbf{l})^{2}}$ is maximal. The reason is that $\widetilde
{I}((1-z)p^{-},\mathbf{k})$ decreases faster than $|\mathbf{k}|^{-1}$ for
$|\mathbf{k}|\gg b_{c}^{-1}$. We get an estimate:
\begin{equation}
|\mathbf{M}_{\text{resc}}|\sim\frac{\pi b_{c}^{-2}}{(2\pi)^{2}}\frac
{1}{|\mathbf{l}|}(2\pi b_{c}^{2})\,|\widetilde{I}(zp^{-},\mathbf{q})|\sim
\frac{1}{|\mathbf{l}|}|\widetilde{I}(zp^{-},\mathbf{q})|\qquad(|\mathbf{q}%
|\gg|\mathbf{l}|\gg b_{c}^{-1})\,\text{,} \label{eq:Mresc}%
\end{equation}
which is comparable to the first term without rescattering in (\ref{eq:Mi}).

\subsection{Impact parameter picture}

In a situation when rescattering cannot be neglected, the separation into
three terms in Eq. (\ref{eq:Mi}) becomes artificial, and we should treat the
whole amplitude as given in (\ref{eq:Mi0}). Substituting the definitions of
$I$, we can transform this expression into a transverse plane integral:%
\begin{equation}
\mathbf{M}_{i}=\frac{i}{2\pi}\int d^{2}\mathbf{y}^{\prime}\,d^{2}%
\mathbf{y}\frac{(\mathbf{y}^{\prime}-\mathbf{y})_{i}}{|\mathbf{y}^{\prime
}-\mathbf{y}|^{2}}e^{-i(\mathbf{q}-\mathbf{l}).\mathbf{y}^{\prime}%
+iz\frac{p^{-}}{2}\Phi(\mathbf{y}^{\prime})}e^{-i\mathbf{l}.\mathbf{y}%
+i(1-z)\frac{p^{-}}{2}\Phi(\mathbf{y})}\,. \label{eq:Miimp}%
\end{equation}
This simple equation provides a key insight into the physics of the process.
Namely, we can view the factor $\Psi(\mathbf{y}^{\prime},\mathbf{y}%
)=\frac{(\mathbf{y}^{\prime}-\mathbf{y})_{i}}{|\mathbf{y}^{\prime}%
-\mathbf{y}|^{2}}$ as the coordinate-space wavefunction of the gluon-quark
state into which quark $B$ splits. Upon crossing the shock wave, this
two-particle wavefunction is multiplied by the eikonal factors $e^{iz\frac
{p^{-}}{2}\Phi(\mathbf{y}^{\prime})}$ and $e^{i(1-z)\frac{p^{-}}{2}%
\Phi(\mathbf{y})}$, depending on the transverse plane position of each
particle. Finally, to compute the S-matrix element, one takes the overlap with
the outgoing state wavefunction $e^{-i(\mathbf{q}-\mathbf{l}).\mathbf{y}%
^{\prime}}e^{-i\mathbf{l}.\mathbf{y}}$.\footnote{See \cite{Kovch1} where
similar considerations are made for a gluon splitting into a $q\bar{q}$ pair
in the field of an incoming nucleus.}

The completely general expression (\ref{eq:Miimp}) can be further simplified
if $|\mathbf{l}|\ll|\mathbf{q}|$. In this case, the typical $\mathbf{y}$
contributing to the integral are much larger than the typical $\mathbf{y}%
^{\prime}$. Approximating $\mathbf{y}^{\prime}-\mathbf{y}\simeq-\mathbf{y}$,
the amplitude takes a factorized form:%
\begin{equation}
\mathbf{M}_{i}\simeq I(zp^{-},q)\,\mathbf{f}_{i}\qquad(|\mathbf{l}%
|\ll|\mathbf{q}|)\,, \label{eq:Miappr}%
\end{equation}
where the gluon emission factor $\mathbf{f}_{i}$ is given by%
\begin{gather}
\mathbf{f}_{i}=-\frac{i}{2\pi}\int d^{2}\mathbf{y}\frac{\mathbf{y}_{i}%
}{|\mathbf{y}|^{2}}e^{-i\mathbf{l}.\mathbf{y}+i(1-z)\left( \frac{b_{c}%
}{|\mathbf{y}|}\right) ^{n}}=-\frac{\mathbf{l}_{i}}{\mathbf{l}^{2}}\times
C_{n}[(1-z)^{1/n}b_{c}|\mathbf{l}|]\,,\label{eq:f}\\[5pt]
C_{n}(u)\equiv\int_{0}^{\infty}dy\,J_{1}(y)\,e^{i(\frac{u}{y})^{n}%
}\,.\nonumber
\end{gather}
This factor has a nontrivial dependence on $\mathbf{l}$. For $|\mathbf{l}|\ll
b_{c}^{-1}$ the integral is dominated by large $|\mathbf{y}|\gg b_{c}^{-1}$,
so that the second term in the exponent can be dropped, giving
\[
\mathbf{f}_{i}\simeq-\mathbf{l}_{i}/\mathbf{l}^{2}\quad(|\mathbf{l}|\ll
b_{c}^{-1})\,.
\]
Thus we recover the first term in Eq. (\ref{eq:Mi}), which in this limit
dominates the amplitude.

On the other hand, in the opposite limit the integral can be evaluated by
stationary phase, with the result:%
\begin{equation}
\mathbf{f}_{i}\simeq\frac{e^{i\times\text{phase}}}{\sqrt{n+1}}\mathbf{l}%
_{i}/\mathbf{l}^{2}\quad(|\mathbf{l}|\gg b_{c}^{-1})\,. \label{eq:largel}%
\end{equation}
Remember that precisely in this case we expected a non-negligible contribution
from rescattering, see Eq.~(\ref{eq:Mresc}). We now see that its effect is
indeed important: rescattering leads to an $\mathcal{O}(1)$ reduction of the
gluon emission amplitude! Physically, this can be explained as follows. To
scatter with large $\mathbf{l}$, the emitted gluon must cross the shock wave
in the region of small impact parameters. In this region, the eikonal factor
in (\ref{eq:f}) distorts the gluon wavefunction, which leads to suppression of
the amplitude via destructive interference. It is however remarkable that, up
to an $\mathbf{l}$-dependent phase, the suppressed amplitude still goes as
$\mathbf{l}_{i}/\mathbf{l}^{2}$.

\begin{figure}[h]
\begin{center}
\includegraphics[
natheight=3.271100in,
natwidth=5.301200in,
height=3.3458in,
width=5.405in
]{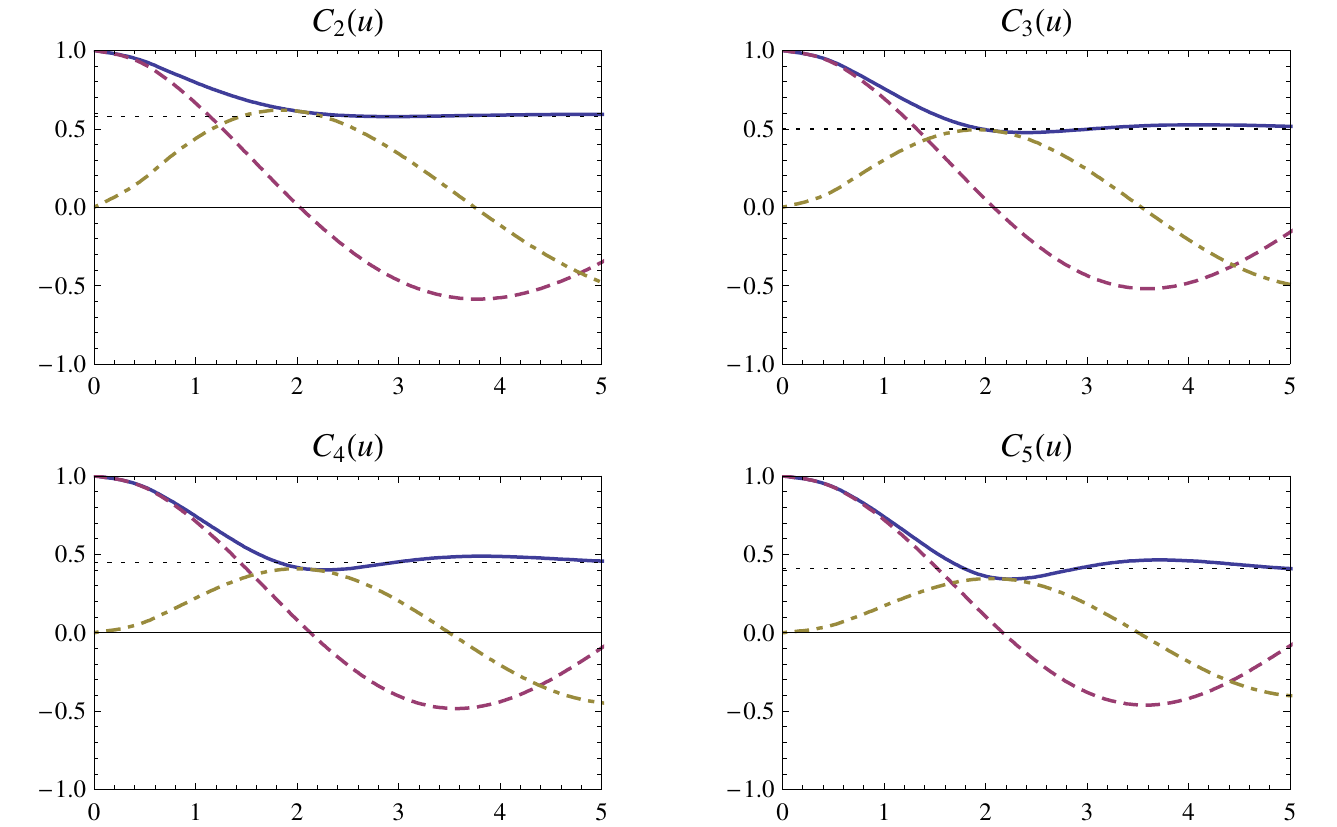}
\end{center}
\caption{\textit{Numerical plots of }$C_{n}(u)$\textit{ for }$n=2,3,4,5$:
\textit{the absolute value (solid blue curve), the real and imaginary parts
(dashed and dot-dashed curves), the }$u=\infty$\textit{ asymptotic value
}$1/\sqrt{n+1}$ \textit{(dotted line).}}%
\label{fig:Cn}%
\end{figure}

In terms of the function $C_{n}$, the above asymptotics can be stated as
follows:%
\[
C_{n}(u)\rightarrow1\quad(u\rightarrow0)\,,\qquad|C_{n}(u)|\simeq1/\sqrt
{n+1}\quad(u\gg1)\,.
\]
As can be seen from Fig.~\ref{fig:Cn}, the large $u$ behavior sets in already
for $u\gtrsim2$.

%EndExpansion

We are now ready to derive the logarithmic correction to the cross section for
$|\mathbf{q}|\gg b_{c}^{-1}$. We square Eq.~(\ref{eq:Miappr}) and integrate in
$|\mathbf{l}|\ll|\mathbf{q}|$, taking into account the suppression for
$|\mathbf{l}|\gg b_{c}^{-1}$. Schematically, we get:
\begin{equation}
\int_{|\mathbf{l}|\lesssim|\mathbf{q}|}d^{2}\mathbf{l}\,|\mathbf{f|}^{2}%
\quad\longrightarrow\quad\int_{\mu_{\text{IR}}}^{b_{c}^{-1}}\frac{dl}{l}%
+\frac{1}{n+1}\int_{b_{c}^{-1}}^{|\mathbf{q}|}\frac{dl}{l}=\log\frac{\mu
_{F}(|\mathbf{q}|)}{\mu_{\text{IR}}}\,. \label{eq:fint}%
\end{equation}
In other words, due to the $1/(n+1)$ suppression of the second term, the
arguments of the two logs combine into a geometric mean which coincides
exactly with Eq.~(\ref{eq:mu})!

The final result is as follows: for $|\mathbf{q}|\gg b_{c}^{-1}$, the NLO
correction to the cross section is given by the same Eq. (\ref{eq:NLO}) as for
$|\mathbf{q}|\lesssim b_{c}^{-1}$ with the following simple replacement:%
\[
\log\frac{\mathbf{q}^{2}}{\mu_{\text{IR}}^{2}}\quad(|\mathbf{q}|\lesssim
b_{c}^{-1})\,\qquad\longrightarrow\qquad\log\frac{\mu_{F}^{2}(|\mathbf{q}%
|)}{\mu_{\text{IR}}^{2}}\quad(|\mathbf{q}|\gg b_{c}^{-1})\,.
\]

\subsection{Discussion}

\label{sec:disc}

Let us discuss what we have achieved. First, we have shown that the emission
amplitude of near-collinear gluons has a factorized form, Eq.~(\ref{eq:Miappr}%
). Physically, it means that we can first consider the hard scattering
process, and worry about adding additional gluons later. If a parton splits in
two before crossing the shock wave, one and only one of the splitting products
absorbs most of the momentum transfer in a typical event. This is actually an
important check of validity of the partonic picture.

Second, we found explicitly the gluon emission factor $\mathbf{f}_{i}$. The
probability distribution of emitted gluons is given by $|\mathbf{f}_{i}|^{2}$.
We found that for large relative transverse momenta $|\mathbf{l}|\gg
b_{c}^{-1}$ (but still $|\mathbf{l}|\ll|\mathbf{q}|$) this distribution is
suppressed by a factor $1/(n+1)$ relative to the standard QCD distribution
$1/\mathbf{l}^{2}$. This is a new effect, which could be used to measure the
number of extra dimensions.

Finally, as a consequence of this suppression, the logarithmic NLO correction
to the partonic cross section involves, for $|\mathbf{q}|\gg b_{c}^{-1}$, a
scale which interpolates between the usual $|\mathbf{q}|$ and $b_{c}^{-1}$ in
agreement with Eq.~(\ref{eq:mu}). In fact, as we show in Appendix
\ref{sec:multi}, such logarithms occur in every order of perturbation theory.
Thus, as usual, they can be exponentiated and removed by shifting the parton
distribution scale to $\mu_{F}(|\mathbf{q}|)$. This, then, provides a formal
justification for the proposal of Emparan, Masip, and Rattazzi
\cite{Emparan:2001kf}, that this scale is the one minimizing higher-order corrections.

Our derivation of the scale (\ref{eq:mu}) has an added advantage that we now
know the gluon distribution. This distribution could not be easily guessed:
for example the standard $1/\mathbf{q}^{2}$ with a sharp $b_{\ast}^{-1}$
cutoff would give rise to the same log. At the same time, the very fact that
we found agreement with Ref.~\cite{Emparan:2001kf} may seem like a mistery.
Remember that they \textit{fixed} this scale to be equal to $b_{\ast}^{-1}$
(see Section \ref{sec:hadronic}). Where, then, does the typical impact
parameter $b_{\ast}$ hide in our computation?

In fact, if we are not interested in the gluon distribution, we can
reformulate our derivation so that it will conform with the original intuition
of \cite{Emparan:2001kf}. The idea is to compute the LHS of Eq. (\ref{eq:fint}%
) not from the asymptotics of $C_{n}$ but directly from the definition of
$\mathbf{f}_{i}$ in the impact parameter representation. Since the $L_{2}$
norm of the gluon wavefunction is the same in the momentum and in the position
space, we have%
\[
\int_{\mu_{\text{IR}}}^{|\mathbf{q}|}\frac{d^{2}\mathbf{l}}{(2\pi)^{2}%
}\,|\mathbf{f}|^{2}\sim\int d^{2}\mathbf{y}\left\vert \frac{\mathbf{y}_{i}%
}{|\mathbf{y}|^{2}}e^{i(1-z)\left( \frac{b_{c}}{|\mathbf{y}|}\right) ^{n}%
}\right\vert ^{2}=\int\frac{d^{2}\mathbf{y}}{\mathbf{y}^{2}}\propto\log
\frac{y_{\max}}{y_{\min}}\,.
\]
The only subtlety is that in the LHS we are not integrating over the whole
$\mathbf{l}$ plane, and thus the limits of the $\mathbf{y}$ intergration have
to be adjusted accordingly. A moment's thought shows that the correct limits
should be put at the typical $\mathbf{y}$ values contributing to
$\mathbf{f}_{i}$ at $|\mathbf{l}|\sim\mu_{\text{IR}}$ and $|\mathbf{l}%
|\sim|\mathbf{q}|$:
\[
y_{\max}\sim\mu_{\text{IR}}^{-1},\quad y_{\max}\sim b_{\ast}\,.
\]
So, we recover the same logarithm as above, and this time $b_{\ast}$ enters explicitly.

We have worked throughout in the toy model of scalar quarks.
However, it should be easy to adapt our considerations to the
realistic case of fermionic matter. One would have to compute the
shock wave crossing vertex for the fermion field. This will require
solving the Dirac equation in the shock wave background. We expect
that our results about factorization and suppression of radiation at
large angles will remain true in the fermionic case as well.

\section{Simultaneous radiation}

\label{sec:simul}

So far we were making the technical simplifying assumption that
particle $A$ does~not QCD-radiate. This is of course not true in
$pp$ collisions at the LHC, when both colliding partons are colored.
We will now discuss how one could relax or remove this restriction.

Consider the following two key properties of QCD radiation:

\begin{enumerate}
\item Near-collinear emission dominates.

\item Its amplitude takes a factorized form.
\end{enumerate}

These properties are true for the standard hard perturbative processes. For
the T-scattering with non-radiating $A$, we have also shown them to be true,
provided that a gluon emission factor is adjusted accordingly. We conjecture
that these properties continue to hold when both $A$ and $B$ are allowed to
radiate. In practice, this implies that the dominant part of the emitted
radiation can be described using the independent emission approximation:~first
compute the radiation off $B$ (taking $A$ classical), then off $A$ (taking $B$ classical).

Intuitively, this can be justified as follows. The fact that the
near-collinear emission dominates is due to the $\sim1/|\mathbf{l}|$
singularity in the gluon emission factor, combined with a phase
space cutoff $|\mathbf{l}|\lesssim|\mathbf{q}|.$ We have seen that
shock wave crossing tends only to suppress radiation at large angles
by distorting the gluon wavefunction, and we do not expect this
tendency to reverse when $A$ is allowed to radiate. Thus collinear
radiation should dominate also for the T-scattering. Once we know
that collinear emission dominates, and thus the emitted radiation
does not change the hard transverse momentum flow of the process, it
seems reasonable that the amplitude should factorize into the
product of the $2\rightarrow2$ scattering with the hardest momentum
exchange times the gluon emission factors.\footnote{We are ignoring
here soft gluons which may be exchanged between particles A and B.
In case of standard hard QCD processes involving two initial
hadrons, like in Drell-Yan, where proofs of factorization to all
orders in perturbation theory exist \cite{Collins}, it is known that
such soft gluon exchanges cancel in the total cross section. We
expect this cancellation to carry over to our case, because soft
gluons do not feel the gravitational field of the energetic
particles (the eikonal factor being proportional to the gluon
energy).}

The independent emission approximation is probably adequate for most practical
purposes. Nevertheless, to try to go beyond it is an interesting theoretical
challenge. We will now describe pictorially a generalization of our formalism
which, we believe, can describe simultaneous emissions from $A$ and $B$
without any extra approximation (except, of course, large CM energy and small
scattering angle).

\begin{figure}[tbh]
\begin{center}
\includegraphics[width=0.50\textwidth]{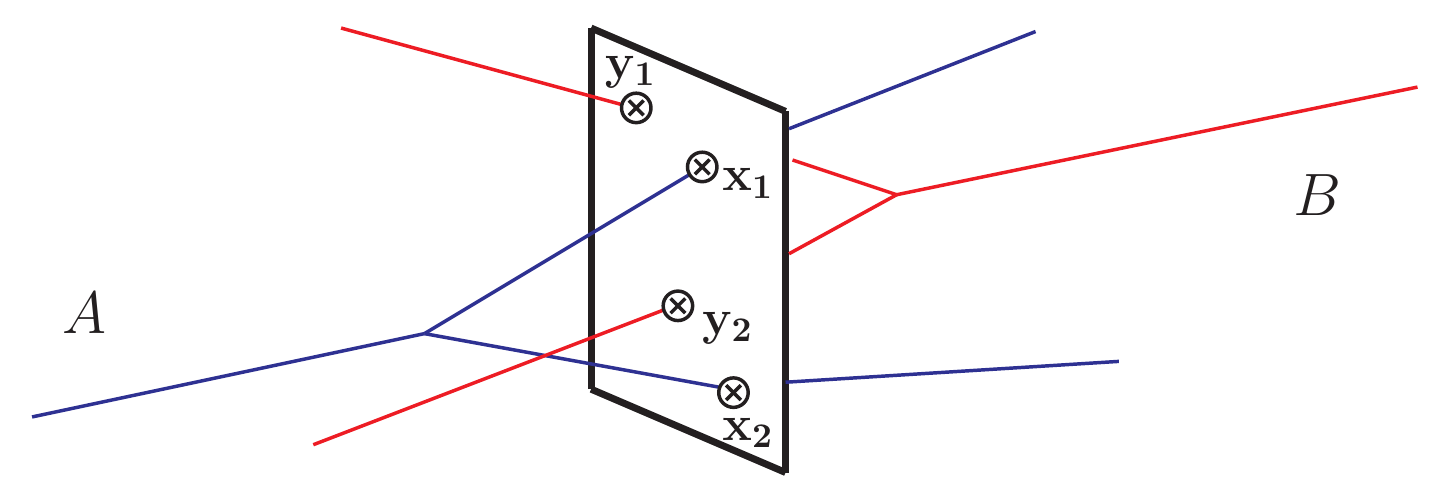}
\end{center}
\caption{\textit{This diagram represents one of the terms in the amplitude
}$A+B\rightarrow A^{\prime}+g+B^{\prime}+g$\textit{, in which both the
right-moving }$A$\textit{ and left--moving }$B$\textit{ split before
colliding. }}%
\label{fig:AB}%
\end{figure}

The starting point is the emission amplitude in the impact parameter
representation~(\ref{eq:Miimp}) (see Appendix \ref{sec:multi} for its
generalizations to two and more gluons). The physical meaning of this equation
in terms of the two-particle wavefunction and individual eikonal factors was
explained above. Suppose now that both particles split. The amplitude can be
constructed according to the following three steps (see Fig.~\ref{fig:AB}).

\begin{enumerate}
\item We evolve the partons from infinity to the transverse plane $x^{-}%
=x^{+}=0$ where the collision is assumed to happen. We introduce many-particle
wavefunctions for the splitting products of both $A$ and $B$. The total
wavefunction in the transverse collision plane is the product of the two:%
\[
\Psi_{tot}=\Psi_{A}(\{\mathbf{x}_{a}\})\Psi_{B}(\{\mathbf{y}_{b}\})\,.
\]
Here $\mathbf{x}_{a}$ $(\mathbf{y}_{b})$ are the transverse coordinates of
left- and right-movers. The $\Psi_{A,B}$ can be computed via the flat-space
light cone perturbation theory. For the one-gluon emission they are the same
as in (\ref{eq:Miimp}).

\item When the splitting products cross the transverse collision plane, the
wavefunction $\Psi_{tot}$ is multiplied with eikonal factors, one for each
pair of opposite-movers. It is not difficult to guess that these factors are
equal to
\[
\exp iz_{a}z_{b}\chi(\mathbf{x}_{a}-\mathbf{y}_{b})\,,
\]
where $\chi$ is the $2\rightarrow2$ eikonal phase from (\ref{eq:eikR}), and
$z_{a}=p_{a}^{+}/p_{A}^{+}$, $z_{b}=p_{b}^{-}/p_{B}^{-}$ are the longitudinal
momentum fractions carried by partons $a$ and $b$.

\item Finally, to find S-matrix elements, we compute the overlap with the
outgoing state wavefunctions. These are simple plane waves if the partons do
not undergo any splittings after the shock wave crossings. If such splittings
are present, like in part (I) of (\ref{eq:1g}), these are more complicated
functions of external momenta. However, since at this stage left- and
right-movers no longer interact, the flat-space light cone perturbation theory
can be used to find them.
\end{enumerate}

We think that this generalized formalism, apart from providing an attractive
mental picture, could find interesting future applications, especially in the
problem of gravitational radiation emission.

\section{Conclusions}

The main point of this paper is that including radiation in small-angle
transplanckian scattering is, after all, a tractable problem. The
gravitational interaction producing eikonal phase factors happens
instantaneously, while the processes of particle splitting are spread in time.
Based on this observation, we developed a formalism which allows explicit
computations of scattering amplitudes in presence of hard, quantum, radiation.
In the impact parameter representation, the radiating particle splits into a
multi-particle virtual state, whose wavefunction, computed via light-cone
perturbation theory, is then multiplied by individual eikonal factors.

We demonstrated the usefulness of our formalism on the concrete problem of
initial state QCD radiation in transplanckian scattering. We will not repeat
here the detailed discussion of the obtained results given in the Introduction
and in Section \ref{sec:disc}. We believe however that this example by no
means exhausts the list of possible applications. We are particularly hopeful
about a possibility to shed new light on the problem of gravitational
radiation emission, which is always on our mind.

\section{Acknowledgements}

We thank Babis Anastasiou, Stefano Frixione, Paolo Nason, Carlo Oleari, and
especially Zoltan Kunszt and Riccardo Rattazzi for useful discussions and
comments. This work is partially supported by the EU under RTN contract
MRTN-CT-2004-503369 and by MIUR under the contract PRIN-2006022501. We thank
the Galileo Galilei Institute for Theoretical Physics for the hospitality and
the INFN for partial support during the completion of this work.

\appendix

\section{Gluon field in the shock wave background}

\label{sec:gluonAS}

Here we will discuss quantization of the free gluon field in the shock wave
background, as well as the gluon emission vertices, the gauge invariance, and
the absence of gluon emission terms localized on the shock wave.

We start from the Maxwell Lagrangian in curved spacetime, which integrating by
parts can be rewritten as
\[
\mathcal{L}=-\frac{1}{4}\sqrt{g}F^{\mu\nu}F_{\lambda\sigma}~\rightarrow
~-\frac{1}{2}\sqrt{g}[(D_{\mu}A_{\nu})(D^{\mu}A^{\nu})-(D_{\mu}A^{\mu}%
)^{2}+^{(4)}\!\!R_{\mu\nu}A^{\mu}A^{\nu}]\,.
\]
Notice that while the $D$-dimensional Ricci tensor of the AS metric is zero
for $x_{\perp}\neq0$ as a consequence of Einstein's equation, the
4-dimensional Ricci tensor of the metric $g_{\mu\nu}$ induced on the SM brane
is nonzero. Namely, it has a nonzero component%
\[
^{(4)}\!R_{--}=-\frac{1}{2}\delta(x^{-})\partial_{\mathbf{x}}^{2}\Phi\neq0
\]
(compare with (\ref{eq:lapl}) where the Laplacian is with respect to all $D-2$
transverse directions).

Let us fix the curved space Lorenz gauge $D_{\mu}A^{\mu}=0.$ In this gauge the
EOM\ take the form:%
\begin{equation}
D^{2}A_{\nu}-^{(4)}\!\!R_{\nu}^{~\mu}A_{\mu}=0\,. \label{eq:gauge}%
\end{equation}

We first discuss the gluon propagator across the shock wave. To find it we
need to solve the EOM with the initial conditions $A_{\mu}=\varepsilon_{\mu
}e^{ipx}$ for $x^{-}<0$.

Consider first the light-cone gauge case $\varepsilon_{+}=0.$ It is easy to
see that the shock wave metric has $\Gamma_{+\nu}^{\mu}\equiv0$, which implies
$A_{+}\equiv0$. The other gauge field components are found to satisfy the
equations:%
\begin{align*}
\partial_{+}\partial_{-}\mathbf{A}+\delta(x^{-})\Phi(\mathbf{x})\partial
_{+}^{2}\mathbf{A} & =\frac{1}{4}\partial_{\mathbf{x}}^{2}\mathbf{A\,,}\\
\partial_{+}\partial_{-}A_{-}+\delta(x^{-})[\Phi(\mathbf{x})\partial_{+}%
^{2}A_{-}+\frac{1}{2}\partial_{i}\Phi(\mathbf{x})\,\partial_{+}\mathbf{A}%
_{i}] & =\frac{1}{4}\partial_{\mathbf{x}}^{2}A_{-}\,,
\end{align*}
Just as in the scalar field case, it is enough to find a matching condition,
i.e.~to relate solution of these equations for $x^{-}=-\epsilon$ and
$x^{-}=+\epsilon.$ One could use the method of Section \ref{sec:tHooft} based
on transforming to the $\widetilde{x}$ coordinates. Here we want to
demonstrate a different, equivalent, approach. Namely, we regularize the above
equations by smearing the $\delta$-functions. One can show that the terms put
in the RHS of the equations can be dropped in this analysis, since their
effect goes to zero when the regulator is removed. All the other terms are
however important.

From the first equation, we find how the transverse components varies across
the shock wave:%
\begin{equation}
\mathbf{A}=\boldsymbol{\upepsilon\,}e^{-i\frac{p^{-}}{2}x^{+}+i\mathbf{p}%
.\mathbf{x}}\exp(i\frac{p^{-}}{2}\int_{-\epsilon}^{x^{-}}\delta(x^{-}%
)\Phi(\mathbf{x}))\,. \label{eq:Ai}%
\end{equation}
Subsituting this solution into the equation for $A_{-}$ we find:%
\begin{equation}
A_{-}=[\varepsilon_{-}-\frac{1}{2}\theta(x^{-})\boldsymbol{\upepsilon}%
_{i}\partial_{i}\Phi(\mathbf{x})]\,e^{-i\frac{p^{-}}{2}x^{+}+i\mathbf{p}%
.\mathbf{x}}\exp(i\frac{p^{-}}{2}\int_{-\epsilon}^{x^{-}}\delta(x^{-}%
)\Phi(\mathbf{x}))\,. \label{eq:A-}%
\end{equation}
Thus on the other side of the shock wave:%
\begin{align*}
\mathbf{A}(x^{-} & =+\epsilon)=\boldsymbol{\upepsilon\,}e^{-i\frac{p^{-}}%
{2}(x^{+}-\Phi(\mathbf{x}))+i\mathbf{p}.\mathbf{x}}\,,\\
A_{-}(x^{-} & =+\epsilon)=(\varepsilon_{-}-\frac{1}{2}%
\boldsymbol{\upepsilon}_{i}\partial_{i}\Phi(\mathbf{x}))e^{-i\frac{p^{-}}%
{2}(x^{+}-\Phi(\mathbf{x}))+i\mathbf{p}.\mathbf{x}}\,.
\end{align*}
These are the desired matching conditions. Taking the Fourier transform we
recover the rule (\ref{eq:gluoncross}).

A general solution of the EOM (\ref{eq:gauge}) is a linear combination of the
just found solution in the gauge $A_{+}=0$ with a pure gauge solution%
\begin{equation}
A_{\mu}=\partial_{\mu}\psi\,. \label{eq:pure}%
\end{equation}
The gauge parameter $\psi$ is given by Eq.~(\ref{eq:modes}) as a general
solution to the Klein-Gordon equation.

Let us now discuss the gluon emission term (\ref{eq:gluonEm}) in a
general gauge. There are several quantities in~(\ref{eq:gluonEm})
with $\delta $-function singularities on the shock wave. Thus one
may wonder if there is a contact emission term localized on the
shock wave. In fact such a contribution is absent, so that one can
always compute the gluon emission as a sum of two separate integrals
for $x^{-}>0$ and $x^{-}<0$. To see this, one can argue as follows.

First of all, as already mentioned in Section \ref{sec:gauge}, localized terms
are absent in the light-cone gauge $A_{+}=0$. In this gauge $A_{\mu}$ does not
contain $\delta$-function singularities as one can see from the explicit
solution (\ref{eq:Ai}),(\ref{eq:A-}). The $\delta$-function does appear in
$g^{++}$ and in $\phi^{\ast}\overleftrightarrow{\partial}_{-}\phi$, but all
contractions involving these terms necessarily contain $A_{+}$ and vanish.

Second, consider the pure gauge case (\ref{eq:pure}). In this case there are
several $\delta$-function terms in (\ref{eq:gluonEm}). However, one can show
that they cancel among themselves. The reason for this cancellation is as
follows. Since the integrand in (\ref{eq:gluonEm}) is a Lorentz invariant, we
can compute it in the $\widetilde{x}$ coordinates (\ref{eq:xt}). In these
coordinates both $\phi$ and $\psi$ are continuous, and the integrand has at
most $\theta$-function singularity on the shock wave.

To demonstrate the absence of localized terms by a concrete example, let us
show that the one-gluon emission amplitude is gauge invariant. We thus have to
show that the amplitude to emit a longitudinally polarized gluon is zero,
without inclusion of any extra terms localized on the shock wave. The
one-gluon emission amplitude is given by the coordinate-space integral:%
\[
\mathcal{M}=i\int d^{4}x\sqrt{g}g^{\mu\nu}\left\{ [\phi_{p^{\prime}%
}^{\text{out}}(x)]^{\ast}\overleftrightarrow{\partial}_{\mu}\phi
_{p}^{\text{in}}(x)\right\} A_{\nu}^{\text{out}}(l,\varepsilon;x)\,.
\]
Here\ $A_{\nu}^{\text{out}}(l,\varepsilon;x)$ is the outgoing gluon
wavefunction. In the considered longitudinal case $\varepsilon_{\mu}=l_{\mu}$
we have:%
\[
A_{\mu}^{\text{out}}(l,\varepsilon;x)\propto\partial_{\mu}\phi_{l}%
^{\text{out}}(x)\,.
\]
The integral splits into two parts: $x^{-}>0$, $x^{-}<0$. Each of these can be
integrated by parts and, using the current conservation, reduces to a boundary
term localized on the shock wave. These boundary terms are not quite
identical:%
\begin{align*}
\mathcal{M}_{(x^{-}>0)} & =-\int dx^{+}d^{2}\mathbf{x}\,\left\{ e^{i\left(
\frac{1}{2}p^{\prime-}x^{+}-\mathbf{p}^{\prime}.\mathbf{x}\right)
}\overleftrightarrow{\partial}_{+}e^{-i\left( \frac{1}{2}p^{-}[x^{+}%
-\Phi(\mathbf{x})]-\mathbf{p}.\mathbf{x}\right) }\right\} e^{i\left(
\frac{1}{2}l^{-}x^{+}-\mathbf{l}.\mathbf{x}\right) }\,,\\
\mathcal{M}_{(x^{-}<0)} & =\int dx^{+}d^{2}\mathbf{x}\,\left\{ e^{i\left(
\frac{1}{2}p^{\prime-}[x^{+}+\Phi(\mathbf{x})]-\mathbf{p}^{\prime}%
.\mathbf{x}\right) }\overleftrightarrow{\partial}_{+}e^{-i\left( \frac{1}%
{2}p^{-}x^{+}-\mathbf{p}.\mathbf{x}\right) }\right\} e^{i\left( \frac{1}%
{2}l^{-}[x^{+}+\Phi(\mathbf{x})]-\mathbf{l}.\mathbf{x}\right) }\,.
\end{align*}
However, after integrating in $x^{+}$ and taking into account the resulting
$p^{-}$-conserving $\delta$-function, they are seen to cancel.

\section{Light-cone phase space}

\label{sec:phase}

The partonic cross section with $n$ gluons emitted $A+B\rightarrow A^{\prime
}+B^{\prime}+g_{1}+\ldots+g_{n}$ is given by the phase space integral
\[
d\hat{\sigma}=\frac{1}{2\hat{s}}|\mathcal{M}_{\text{rel}}^{(n)}|^{2}%
\,(2\pi)^{4}\delta^{(4)}(p_{i}-p_{f})\,d\Phi_{(n+2)}\,.
\]
We assume that $A$ and $B$ collide head on along the $z$ direction. We use the
light-cone phase space adapted to the direction of motion of each particle:%
\begin{gather*}
d\Phi_{(n+2)}=\frac{dp_{A^{\prime}}^{+}d^{2}\mathbf{p}_{A^{\prime}}}%
{(2\pi)^{3}2p_{A^{\prime}}^{+}}\frac{dp_{B^{\prime}}^{-}d^{2}\mathbf{p}%
_{B^{\prime}}}{(2\pi)^{3}2p_{B^{\prime}}^{-}}\prod\frac{dl_{i}^{-}%
d^{2}\mathbf{l}_{i}}{(2\pi)^{3}2l_{i}^{-}},\\[5pt]
\delta^{(4)}(p_{i}-p_{f})=2\delta(p_{i}^{+}-p_{f}^{+})\delta(p_{i}^{-}%
-p_{f}^{-})\delta^{(2)}(\mathbf{p}_{i}-\mathbf{p}_{f})\,.
\end{gather*}
The momentum conserving $\delta$-function is saturated by integrating in
$dp_{A^{\prime}}^{+}d^{2}\mathbf{p}_{A^{\prime}}dp_{B^{\prime}}^{-}$, which
gives%
\[
d\hat{\sigma}=\frac{1}{16\pi^{2}\hat{s}^{2}}|\mathcal{M}_{\text{rel}}%
^{(n)}|^{2}\,\frac{p_{B}^{-}}{p_{B^{\prime}}^{-}}d^{2}\mathbf{p}_{B^{\prime}%
}\prod\frac{dl_{i}^{-}d^{2}\mathbf{l}_{i}}{(2\pi)^{3}2l_{i}^{-}}\,.
\]
The differential cross section in the momentum transfer $\mathbf{q}$ and the
Bjorken $x$ is thus given by:%
\[
\frac{d\hat{\sigma}}{d^{2}\mathbf{q\,}dx}=\frac{1}{16\pi^{2}\hat{s}^{2}}%
\int|\mathcal{M}_{\text{rel}}^{(n)}|^{2}\,\frac{p_{B}^{-}}{p_{B^{\prime}}^{-}%
}\prod\frac{dl_{i}^{-}d^{2}\mathbf{l}_{i}}{(2\pi)^{3}2l_{i}^{-}}%
\delta\Bigl(x-\frac{\mathbf{q}^{2}}{p_{B}^{-}q^{+}}\Bigr),
\]
where $q^{+}=\mathbf{p}_{B^{\prime}}^{2}/p_{B^{\prime}}^{-}+\sum\mathbf{l}%
_{i}^{2}/l_{i}^{-},$ $\mathbf{p}_{B^{\prime}}=\mathbf{q}-\sum\mathbf{l}%
_{i},~p_{B^{\prime}}^{-}=p_{B}^{-}-\sum l_{i}^{-}$ under the integral sign.

\section{Multi-gluon emission}

\label{sec:multi}

Let us look at the leading logarithmic corrections to the cross section which
come from the radiation of many gluons. We want to verify that they have the
correct form in order to be reabsorbed into the PDFs normalized at the scale
$\mu_{F}(q)$.

\begin{figure}[tbh]
\begin{center}%
\begin{tabular}
[c]{cc}%
\includegraphics[width=0.46\textwidth]{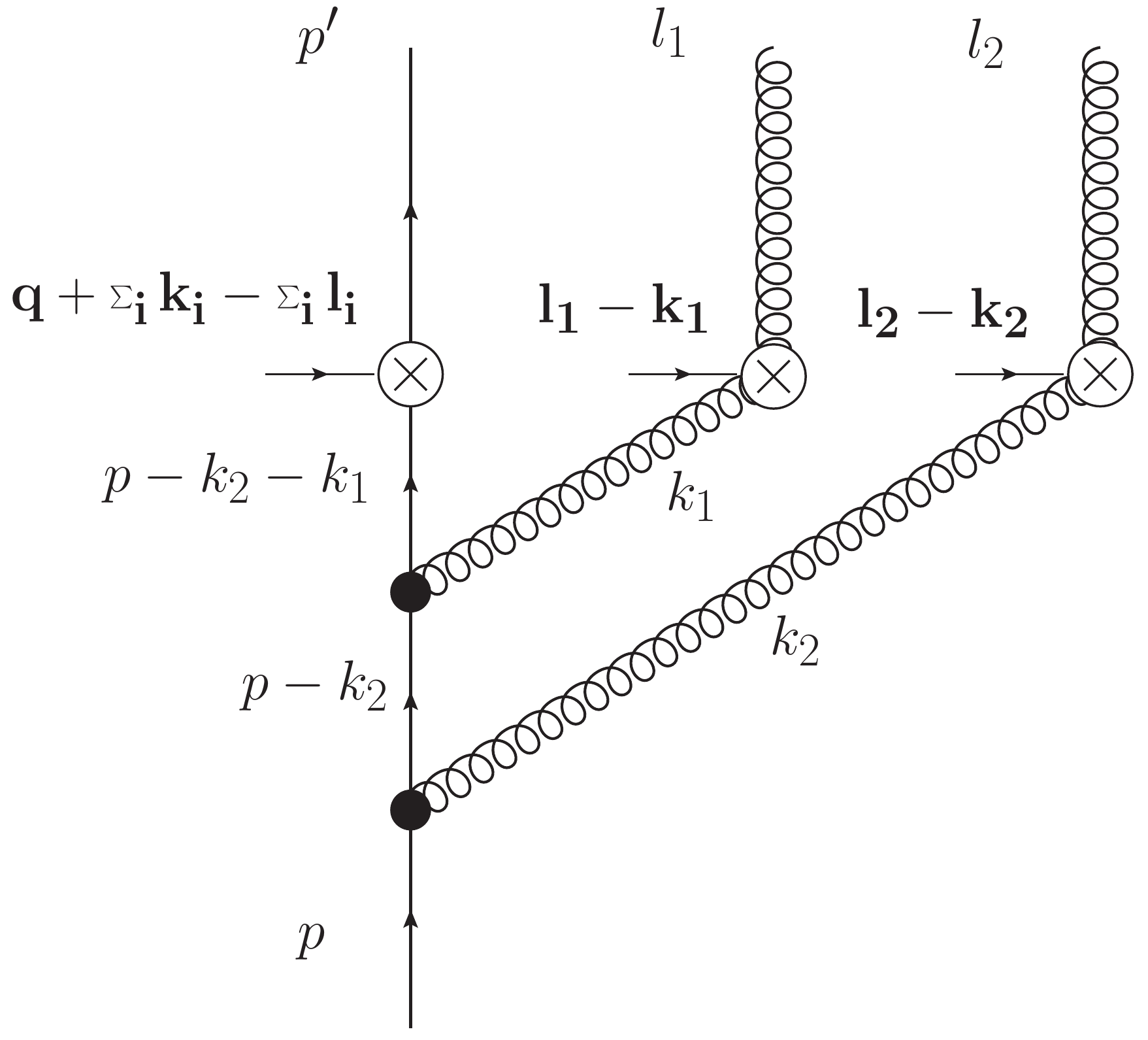} &
\includegraphics[width=0.46\textwidth]{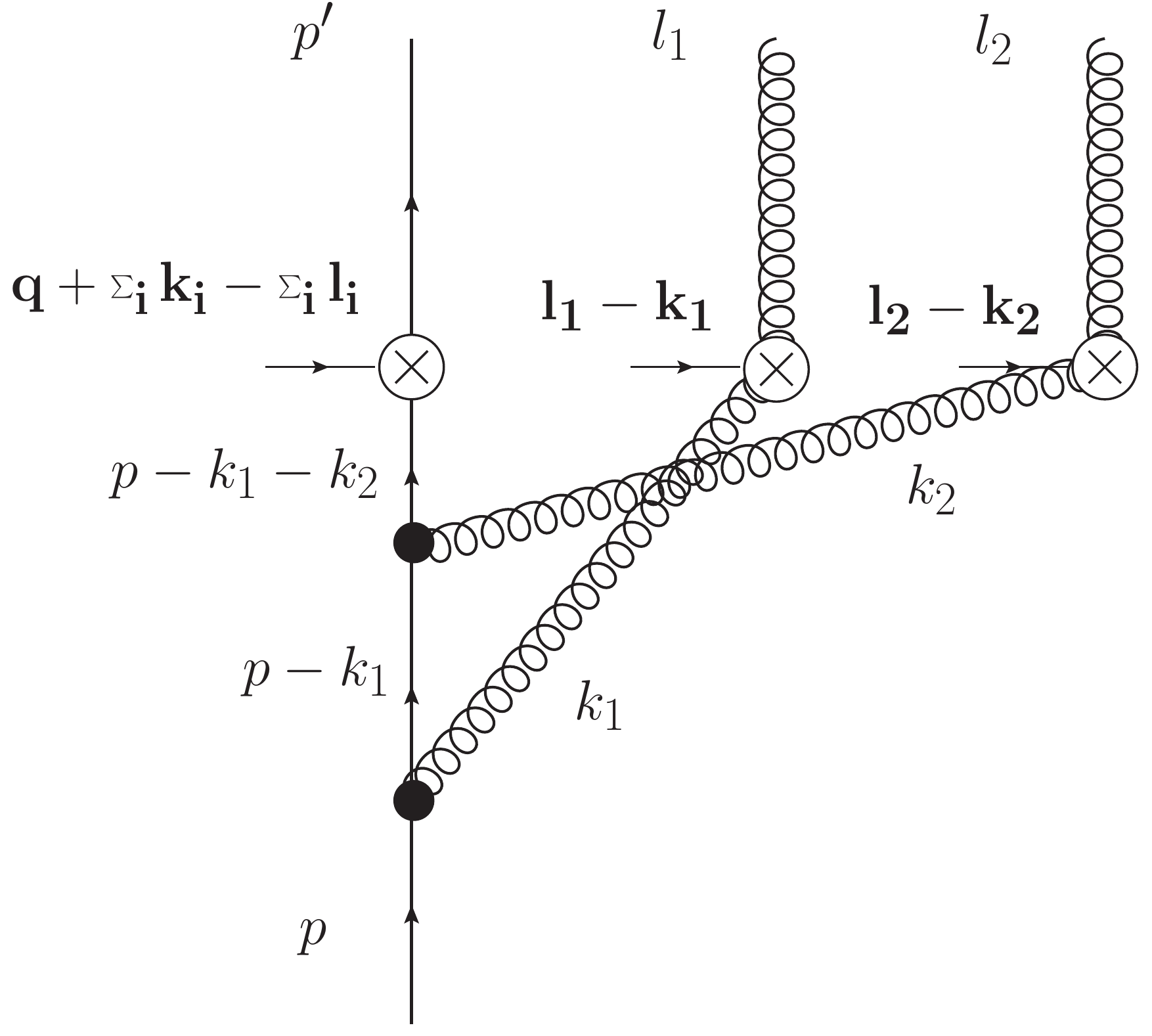}
\end{tabular}
\end{center}
\caption{\textit{Relevant diagrams for the emission of 2 gluons.}}%
\label{figure2gluons}%
\end{figure}

To begin with, consider the emission of two gluons. In the LLA, we are looking
for $(\alpha_{s}\log\frac{\mu_{F}}{\mu_{\text{IR}}})^{2}$ corrections to the
cross section. One can show that, because of the form of the denominators in
our Feynman rules, a gluon which does not cross the shock wave will not give
rise to a large logarithm. Thus the relevant diagrams are those with both
gluons emitted at $x^{-}<0$, shown in Fig.~\ref{figure2gluons}. Let us
consider the first of these diagrams, and define:
\[
\left\{
\begin{array}
[c]{l}%
p^{\prime-}=z_{1}z_{2}p^{-}\\
l_{2}^{-}=(1-z_{2})p^{-}\\
l_{1}^{-}=z_{2}(1-z_{1})p^{-}\,.
\end{array}
\right.
\]
Computing this amplitude using the Feynman rules and going into the impact
parameter representation, we find $\mathcal{M}^{(l_{2}l_{1})}=-8p^{\prime
-}(ig_{s})^{2}\,\mathbf{M}_{ij}^{(l_{2}l_{1})}\boldsymbol{\upepsilon_{1}}%
_{i}\boldsymbol{\upepsilon_{2}}_{j}$, with:
\begin{align*}
\mathbf{M}_{ij}^{(l_{2}l_{1})} & =\int\frac{d^{2}\mathbf{k_{1}}}{(2\pi)^{2}%
}\frac{d^{2}\mathbf{k_{2}}}{(2\pi)^{2}}\frac{\mathbf{k_{2}}_{j}}%
{(\mathbf{k_{2}})^{2}}\frac{(\frac{1}{1-z_{1}}\mathbf{k_{1}}+\mathbf{k_{2}%
})_{i}}{(\frac{1}{1-z_{1}}\mathbf{k_{1}}^{2}+\frac{1-z_{2}+z_{1}z_{2}}%
{1-z_{2}}\mathbf{k_{2}}^{2}+2\mathbf{k_{1}}\cdot\mathbf{k_{2}})}\\
& \qquad\times I(z_{1}z_{2}p^{-},\mathbf{q+k_{1}+k_{2}-l_{1}-l_{2}}%
)\,I(z_{2}(1-z_{1})p^{-},\mathbf{l_{1}-k_{1}})\,I((1-z_{2})p^{-}%
,\mathbf{l_{2}-k_{2}})\\
& \equiv-\frac{1}{(2\pi)^{2}}\int d^{2}\mathbf{x}\,d^{2}\mathbf{y_{1}}%
\,d^{2}\mathbf{y_{2}}\frac{(\mathbf{x}-\mathbf{y_{1}})_{i}}{|\mathbf{x}%
-\mathbf{y_{1}}|^{2}}\frac{(\mathbf{x}-\mathbf{y_{2}})_{j}+B(\mathbf{x}%
-\mathbf{y_{1}})_{j}}{|(\mathbf{x}-\mathbf{y_{2}})+B(\mathbf{x}-\mathbf{y_{1}%
})|^{2}+A|\mathbf{x}-\mathbf{y_{1}}|^{2}}\\
& \qquad\times e^{-i(\mathbf{q}-\mathbf{l_{1}}-\mathbf{l_{2}}).\mathbf{x}%
+iz_{1}z_{2}\frac{p^{-}}{2}\Phi(\mathbf{x})}e^{-i\mathbf{l_{1}}.\mathbf{y_{1}%
}+iz_{1}(1-z_{2})\frac{p^{-}}{2}\Phi(\mathbf{y_{1}})}e^{-i\mathbf{l_{2}%
}.\mathbf{y_{2}}+i(1-z_{2})\frac{p^{-}}{2}\Phi(\mathbf{y_{2}})}\,\,,
\end{align*}
where $A=z_{1}(1-z_{1})/(1-z_{2})$, $B=1-z_{1}.$ The integrand has the
expected form (three-particle wavefunction in the transverse plane)$\times
$(individual eikonal factors)$\times$(outgoing states).

Let us estimate this amplitude for $|\mathbf{q}|\gg|\mathbf{l_{1}%
}|,|\mathbf{l_{2}}|$. In this limit we will find the double logarithm
associated with $P_{Q\rightarrow Q}(z_{1})P_{Q\rightarrow Q}(z_{2})$, while in
the other regions there are those associated with splittings into gluons. Now,
the integral in $\mathbf{x}$ is dominated by values much smaller than those
dominating the integrals in $\mathbf{y_{1}}$ and $\mathbf{y_{2}}$. Neglecting
$\mathbf{x}$ with respect to $\mathbf{y_{1}}$ and $\mathbf{y_{2}}$, we
obtain:
\begin{align}
\mathbf{M}_{ij}^{(l_{2}l_{1})} & \simeq I(z_{1}z_{2}p^{-},\mathbf{q}%
)\,\,\frac{1}{(2\pi)^{2}}\int d^{2}\mathbf{y_{1}}\frac{\mathbf{y_{1}}_{i}%
}{(\mathbf{y_{1}})^{2}}e^{-i\mathbf{l_{1}y_{1}}}e^{iz_{2}(1-z_{1})\left(
\frac{b_{c}}{|\mathbf{y_{1}}|}\right) ^{n}}\nonumber\\
& \qquad\times\frac{1}{(2\pi)^{2}}\int d^{2}\mathbf{y_{2}}\frac
{\mathbf{y_{2}}_{j}+B\mathbf{y_{1}}_{j}}{(\mathbf{y_{2}}+B\mathbf{y_{1}}%
)^{2}+A\mathbf{y_{1}}^{2}}e^{-i\mathbf{l_{2}y_{2}}}e^{i(1-z_{2})\left(
\frac{b_{c}}{|\mathbf{y_{2}}|}\right) ^{n}}\,\,.
\label{2gluonsimpactparspace}%
\end{align}

A double logarithm can be obtained only if the amplitude behaves like
$|\mathbf{M}_{ij}|\sim\frac{1}{|\mathbf{l_{1}}|}\frac{1}{|\mathbf{l_{2}}|}$.
In turn, this means that we must look for the situation in which the integrand
behaves like $\frac{\mathbf{y_{1}}_{i}}{|\mathbf{y_{1}}|^{2}}\frac
{\mathbf{y_{2}}_{j}}{|\mathbf{y_{2}}|^{2}}$, which is true only if the
$\mathbf{y_{1}}$ integral is dominated by much smaller values than those
dominating the $\mathbf{y_{2}}$ integral. This will happen if and only if
$|\mathbf{l_{1}}|\gg|\mathbf{l_{2}}|$. We conclude that the gluon emitted
closer to the shock wave must have a larger transverse momentum. We will come
back to this later, when we generalize to an arbitrary number of gluons.

Thus for $|\mathbf{l_{1}}|\gg|\mathbf{l_{2}}|$, we can neglect $\mathbf{y_{1}%
}$ with respect to $\mathbf{y_{2}}$ in the integrand, and the last integral
factorizes giving:
\[
\mathbf{M}_{ij}^{(l_{2}l_{1})}\simeq I(z_{1}z_{2}p^{-},\mathbf{q}%
)\,\,\mathbf{f}_{i}(l_{1}^{-}\text{,}\mathbf{l_{1}})\,\,\mathbf{f}_{j}%
(l_{2}^{-}\text{,}\mathbf{l_{2}})\qquad(|\mathbf{l_{1}}|\gg|\mathbf{l_{2}%
}|)\,,
\]
where $\mathbf{f}_{i}$ is the gluon emission factor introduced in
Eq.~(\ref{eq:f}). After squaring and integrating in $\mathbf{l_{1}}$ and
$\mathbf{l_{2}}$, this gives a double logarithm of $\mu_{F}(q)$.

For completeness let us verify explicitly that the contribution from the
region $|\mathbf{l_{1}}|\ll|\mathbf{l_{2}}|$ is subdominant. One can show that
the amplitude depends on $\mathbf{l_{1}}$ and $\mathbf{l_{2}}$ as follows:
\[
\mathbf{M}_{ij}^{(l_{2}l_{1})}\propto\left\{
\begin{array}
[c]{lr}%
\frac{\mathbf{l_{2}}_{j}\mathbf{l_{2}}_{i}}{|\mathbf{l_{2}}|^{4}} &
\mbox{ if }\,|\mathbf{l_{1}}|\ll|\mathbf{l_{2}}|\ll b_{c}^{-1}\\
b_{c}^{\frac{n}{n+1}}\left( \frac{1}{|\mathbf{l_{2}}|}\right) ^{1+\frac
{1}{n+1}}\,\log\left( \frac{|\mathbf{l_{2}}|^{\frac{1}{n+1}}}{|\mathbf{l_{1}%
}|\,b_{c}^{\frac{n}{n+1}}}\right) & \mbox{ if }\,|\mathbf{l_{1}}|\ll
b_{c}^{-1}\ll|\mathbf{l_{2}}|\\
\frac{\mathbf{l_{1}}_{i}\mathbf{l_{1}}_{j}}{|\mathbf{l_{1}}|^{4}}\,\left(
\frac{|\mathbf{l_{1}}|}{|\mathbf{l_{2}}|}\right) ^{\frac{1}{n+1}} &
\mbox{ if }\,b_{c}^{-1}\ll|\mathbf{l_{1}}|\ll|\mathbf{l_{2}}|
\end{array}
\quad.\right.
\]
It is immediate to see that from each of the three regions we only get single logarithms.

Thus, the region $|\mathbf{l_{2}}|<|\mathbf{l_{1}}|<|\mathbf{q}|$ dominates,
and in the LLA we get:
\[
\int\int d^{2}\mathbf{l}_{1}d^{2}\mathbf{l}_{2}|\mathbf{M}_{ij}^{(l_{2}l_{1}%
)}|^{2}\approx\frac{1}{2}|{I}(z_{1}z_{2}p^{-},\mathbf{q})|^{2}\left\{ \pi
\log\frac{\mu_{F}(q)^{2}}{\mu^{2}}\right\} ^{2},
\]
where the $\frac{1}{2}$ factor comes from the $\theta(|\mathbf{l}%
_{1}|-|\mathbf{l}_{2}|)$.

The second diagram in Fig.~\ref{figure2gluons} is the same with $l_{1}%
\leftrightarrow l_{2}$. Since the leading contributions come from integration
over different regions of transverse momenta, in the LLA there is no
interference between the two diagrams. Putting all together and including also
the double logarithms arising from the regions $|\mathbf{q}|\gg|\mathbf{q}%
-\mathbf{l}_{1}|,|\mathbf{l}_{2}|$ and $|\mathbf{q}|\gg|\mathbf{l}%
_{1}|,|\mathbf{q}-\mathbf{l}_{2}|$, which can be computed in a similar way, we
finally obtain:
\begin{align*}
\frac{d\hat{\sigma}_{\text{NLO}}}{d^{2}\mathbf{q}\,dz_{1}dz_{2}} &
\simeq\frac{1}{4\pi^{2}}\left\{ \bigl|\widetilde{I}(z_{1}z_{2}p^{-}%
,\mathbf{q})\bigr|^{2}P_{Q\rightarrow Q}(z_{1})P_{Q\rightarrow Q}%
(z_{2})+\bigl|\widetilde{I}((1-z_{2})p^{-},\mathbf{q})\bigr|^{2}%
P_{Q\rightarrow g}(1-z_{2})P_{Q\rightarrow Q}(z_{1})\right. \\
& \qquad+\left. \bigl|\widetilde{I}(z_{2}(1-z_{1})p^{-},\mathbf{q}%
)\bigr|^{2}P_{Q\rightarrow g}(1-z_{1})P_{Q\rightarrow Q}(z_{2})\right\}
\frac{1}{2!}\left[ \frac{\alpha_{s}}{2\pi}\log\frac{\mu_{F}^{2}(q)}{\mu^{2}%
}\right] ^{2}.
\end{align*}
which shows that the optimal factorization scale is $\mu_{F}(q)$.

\begin{figure}[tbh]
\begin{center}
\includegraphics[width=0.60\textwidth]{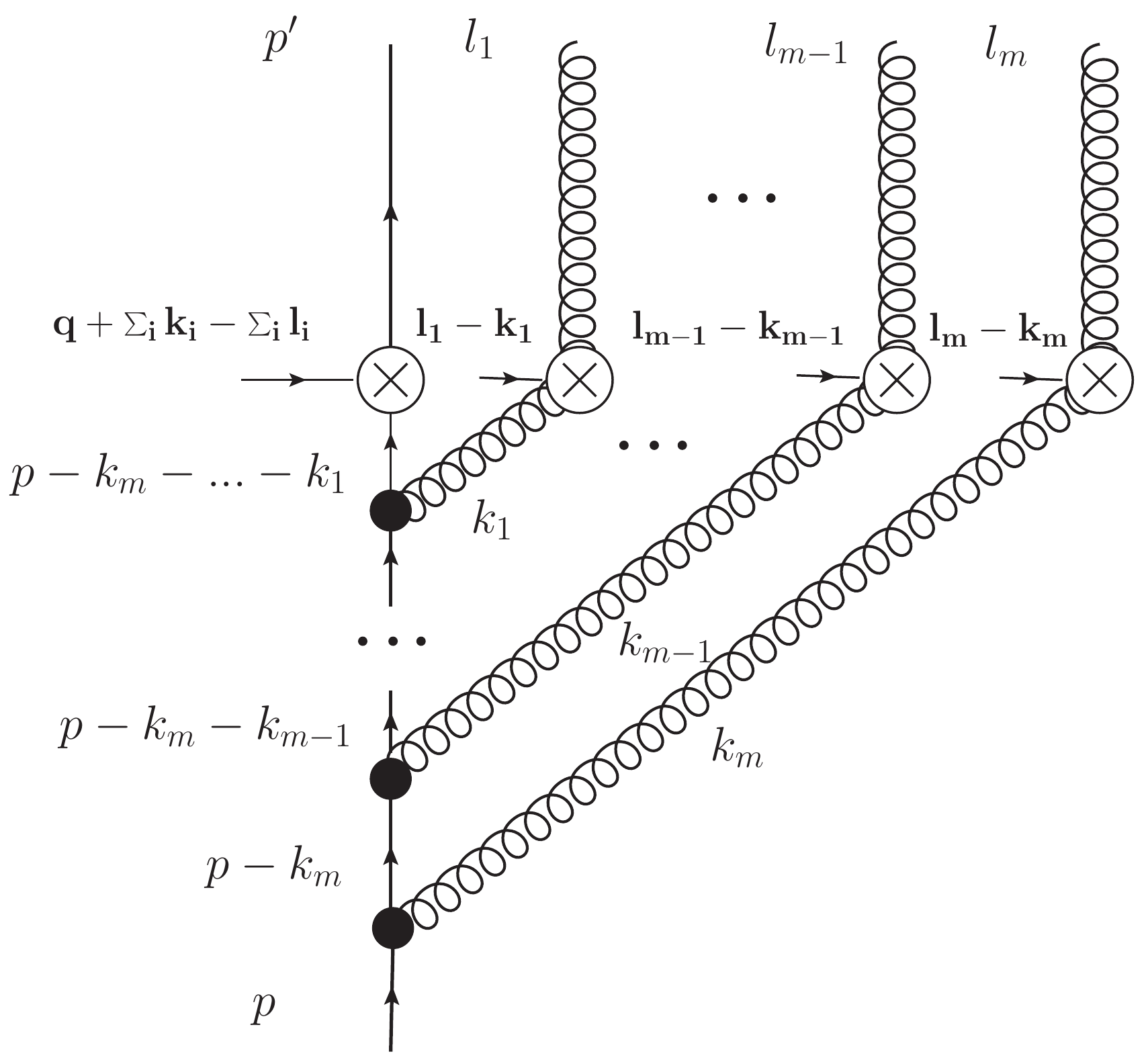}
\end{center}
\caption{\textit{Typical relevant diagram for the emission of }$m$\textit{
gluons.}}%
\label{figureMgluons}%
\end{figure}

Generalization to an arbitrary number $m$ of emitted gluons is
straightforward. Consider the typical relevant diagram shown in
Fig.~\ref{figureMgluons}, and let us look at the transverse momenta involved
in the quark-gluon vertices. In the first one from below only $\mathbf{k_{m}}$
appears, thus the contribution to the amplitude will be $\frac{\mathbf{k_{m}%
}\cdot\,\boldsymbol{\upepsilon_{m}}}{|\mathbf{k_{m}}|^{2}}$, like in the case
of one gluon emission. In the next vertex both $\mathbf{k_{m}}$ and
$\mathbf{k_{m-1}}$ are involved, but it is obvious that its contribution
should reduce to $\frac{\mathbf{k_{m-1}}\cdot\,\boldsymbol{\upepsilon_{m-1}}%
}{|\mathbf{k_{m-1}}|^{2}}$ when $|\mathbf{k_{m}}|\ll|\mathbf{k_{m-1}}|$. This
is true for all the $m$ vertices: if the transverse momentum $\mathbf{k}$ of a
gluon is much larger than those of the gluons which were emitted previously,
then its contribution to the amplitude will be $\frac{\mathbf{k}%
\cdot\,\boldsymbol{\upepsilon}}{|\mathbf{k}|^{2}}$. Thus in general we can
write, for the diagram in Fig.~\ref{figureMgluons}:
\[
M^{(l_{m}\,...\,l_{1})}=-2^{m+1}p^{\prime-}(ig_{s})^{m}\,\mathbf{M}%
_{i_{1}\,...\,i_{m}}^{(l_{m}\,...\,l_{1})}\prod_{j=1}^{m}%
(\boldsymbol{\upepsilon}_{\mathbf{j}})_{i_{j}}\,,
\]
where:
\begin{align*}
\mathbf{M}_{i_{1}\,...\,i_{m}}^{(l_{m}\,...\,l_{1})} & =\int\left(
\prod_{j=1}^{m}\frac{d^{2}\mathbf{k_{j}}}{(2\pi)^{2}}\right) \,\left(
\prod_{j=1}^{m}\frac{(\mathbf{k_{j}}+\sum_{r>j}\mathcal{O}(\mathbf{k_{r}%
}))_{i_{j}}}{\mathbf{k_{j}}^{2}+\sum_{r>j}\mathcal{O}(\mathbf{k_{j}}%
\cdot\mathbf{k_{r}})+\sum_{r,s>j}\mathcal{O}(\mathbf{k_{s}}\cdot\mathbf{k_{r}%
})}\right) \times\\
& \qquad\times I(z_{1}z_{2}p^{-},\mathbf{q+\sum_{j=1}^{m}(k_{j}-l_{j}%
)})\,\times\left( \prod_{j=1}^{m}I(l_{j}^{-},\mathbf{l_{j}-k_{j}})\right)
\,.
\end{align*}
Fourier-transforming this expression and considering the limit $|\mathbf{q}%
|\gg|\mathbf{l_{j}}|\,(j=1,...,m)$, we would obtain an expression analogous to
(\ref{2gluonsimpactparspace}) with $m$ integrals instead of two. From the
structure of the amplitude in $\mathbf{k}$ space, it is clear that it will
be:
\[
\mathbf{M}_{i_{1}\,...\,i_{m}}^{(l_{m}\,...\,l_{1})}\simeq I(z_{1}z_{2}%
p^{-},\mathbf{q})\,\,\frac{1}{(2\pi)^{2m}}\int\left( \prod_{j=1}^{m}%
d^{2}\mathbf{y_{j}\,}e^{-i\mathbf{l_{j}y_{j}}}\,e^{i\frac{l_{j}^{-}}{p^{-}%
}\left( \frac{b_{c}}{|\mathbf{y_{j}}|}\right) ^{n}}\right) \Psi
(\mathbf{y_{1}},...,\mathbf{y_{m}})\,,
\]
where the multi-particle wavefunction $\Psi$:
\[
\Psi(\mathbf{y_{1}},...,\mathbf{y_{m}})\simeq\prod_{j=1}^{m}\frac
{(\mathbf{y_{j}})_{i_{j}}}{(\mathbf{y_{j}})^{2}}\quad\mbox{ if }\quad
|\mathbf{y_{1}}|\ll|\mathbf{y_{2}}|\ll...\ll|\mathbf{y_{m}}|\,.
\]
This means that the only term with $m^{th}$ power of a large logarithm comes
from the region $|\mathbf{l_{1}}|\gg|\mathbf{l_{2}}|\gg...\gg|\mathbf{l_{m}}%
|$. When computing the total cross section with $m$ identical gluons in the
final state, we can always reorder the gluons so that $|\mathbf{l_{1}%
}|>|\mathbf{l_{2}}|>...>|\mathbf{l_{m}}|$. Then in the LLA only the shown
diagram contributes, with a $\frac{1}{m!}$ factor from $\theta
(|\mathbf{l_{close}}|>...>|\mathbf{l_{far}}|).$ Putting all together, we get
\[
\frac{d\hat{\sigma}_{\text{NLO}}}{d^{2}\mathbf{q}}\simeq\frac{1}{4\pi^{2}%
}\bigl|\widetilde{I}(p^{-}\prod_{j=1}^{m}z_{j},\mathbf{q})\bigr|^{2}%
\,\prod_{j=1}^{m}P_{Q\rightarrow Q}(z_{j})\,dz_{j}\,\frac{1}{m!}\left[
\frac{\alpha_{s}}{2\pi}\log\frac{\mu_{F}^{2}(q)}{\mu^{2}}\right] ^{m},
\]
where:
\[
z_{j}=1-\sum_{i=j}^{m}l_{i}^{-}/p^{-}.
\]
All these large logarithms are absorbed into the PDF normalized at the scale
$\mu_{F}(|\mathbf{q}|)$.

\end{document}